\documentclass[a4paper,fleqn,usenatbib,useAMS]{mnras}

\usepackage{graphicx}   
\usepackage{amsmath}    
\usepackage{amssymb}    
\usepackage{multicol}        
\usepackage{bm}         
\usepackage{pdflscape}  

\usepackage{xcolor}

\usepackage[T1]{fontenc}
\usepackage{ae,aecompl}

\title[Auriga]{Large scale star formation in Auriga region}

\author[XYZ]{
A. K. Pandey,$^{1}$\thanks{E-mail: pandey@aries.res.in}
Saurabh Sharma,$^{1}$, N. Kobayashi,$^{2,3}$ 
\newauthor
~Y. Sarugaku,$^{3}$ and K. Ogura$^{4}$
\\\\
$^{1}$Aryabhatta Research Institute of Observational Sciences (ARIES), Manora Peak, Nainital, 263 002, India\\
$^{2}$Institute of Astronomy, University of Tokyo, 2-21-1 Osawa, Mitaka, Tokyo 181-0015, Japan\\
$^{3}$Kiso Observatory, School of Science, University of Tokyo, Mitake-mura, Kiso-gun, Nagano 397-0101, Japan\\
$^{4}$Kokugakuin University, Higashi, Shibuya-ku, Tokyo 150-8440, Japan\\
}

\date{Accepted XXX. Received YYY; in original form ZZZ}

\pubyear{2019}

\begin{document}
\label{firstpage}
\pagerange{\pageref{firstpage}--\pageref{lastpage}}
\maketitle

\begin{abstract}
New observations in the $VI$ bands along with archival data from the 2MASS and $WISE$ surveys have been used to generate 
a catalog of young stellar objects (YSOs) covering an area of about  $6^\circ\times6^\circ$ in 
the Auriga region centered at $l\sim173^\circ$ and $b\sim1^\circ.5$.  The nature of the identified YSOs and their 
spatial distribution are used to study the star formation in the region. The distribution of YSOs 
along with that of the ionized and molecular gas reveals two ring-like structures stretching over
 an area of a few degrees each in extent. We name these structures as Auriga Bubbles 1 and 2. 
The center of the Bubbles appears to be above the Galactic mid-plane.
The majority of Class\,{\sc i} YSOs are associated with the Bubbles, whereas the relatively older population, i.e., 
Class\,{\sc ii} objects are rather randomly distributed. Using the minimum spanning tree analysis,
 we found 26 probable sub-clusters having 5 or more members. The sub-clusters are between  $\sim$0.5 pc - $\sim$ 3 pc
 in size and are somewhat elongated. The star formation efficiency in most of the  sub-cluster region
varies between 5$\%$ - 20$\%$ 
indicating that the sub-clusters could be bound regions. The radii of these sub-clusters also support it. 
\end{abstract}



\section{Introduction}

Massive stars (mass $\gtrsim$ 8 M$_\odot$, OB spectral types) are usually found in stellar groups, i.e., star clusters or OB 
associations. This is expected as a majority of the stars are born in groups embedded in molecular clouds 
\citep[see e.g.,][]{2003ARAA..41...57L}. However, the recent study by \citet{2018MNRAS.475.5659W} favors the hierarchical star 
formation model, in which a minority of stars forms in bound clusters and large-scale, hierarchically-structured associations are 
formed in-situ.  Recent mid-infrared (MIR) surveys have also shown that a significant number of young stellar objects (YSOs) form in the 
distributed mode \citep[see e.g.,][]{2008ApJ...688.1142K,2019AJ....157..112P}.

O stars are the major source of Lyman continuum (Lyc) ionizing  radiation. Also strong stellar winds from O and B stars release 
kinematic energy of up to 10$^{51}$ ergs over their lifetimes \citep[][]{1987ApJ...317..190M}.  Since young star clusters/ OB 
associations have several OB stars (mass $\gtrsim$ 8 M$_\odot$), the ionizing radiation and stellar wind of these stars are the 
dominant input power source into the surroundings at the initial phase, creating a shell/ bubble of radius $\leq$ 100 pc at the  end 
of the wind-driven phase \citep{1987ApJ...317..190M}. These shells can be observed in H\,{\sc i} 21 cm line observations, also they 
can be observable in X-ray from their hot interiors, in optical due to emission from ionized gas and in infrared (IR) from swept up dust 
in the shell \citep{2005AJ....129..393O,2014MNRAS.438.1089C,2019MNRAS.484.1800S}. 
Actually expanding shells of dense gas around H\,{\sc ii} regions have been found in several previous studies, e.g., around 
$\lambda$ Orionis, Gemini OB1, W33, the Cepheus Bubble \citep[cf.][and reference therein]{1998ApJ...507..241P}.  In the course 
of expansion of the shells the remaining molecular cloud could be induced to form a second generation of stars.

However the energy input by these massive stars to the surroundings toward the end phase is probably dominated by their supernova 
(SN) explosions, which could release energy of $\sim$ 10$^{51}$ ergs; however, only a few cases are known which provide direct 
evidence of supernova-remnant (SNR)-molecular cloud interaction \citep{1987ApJ...317..190M,1998ApJ...507..241P, 
2006ApJ...649..759C}.  The expansion of the SN blast waves also could produce a new generation of stars. Only morphological  
signatures  are  often  invoked  to prove a physical association between SNRs and star-forming molecular clouds \citep[e.g.,][and 
reference therein]{2001AJ....121..347R}.  The most convincing evidence is the line broadening due to shocked molecular gas 
\citep{1998ApJ...508..690F,1999ApJ...511..836R}.

\citet{2012AJ....143...75K} made H\,{\sc i} 21 cm line observations and detected high velocity H\,{\sc i} gas at  
$l\sim$173$^\circ, b\sim$1.5$^\circ $. They suggested a large scale star formation activity possibly associated with an SN explosion  
in the H\,{\sc ii} complex G173+1.5. The low resolution H\,{\sc i} survey reveals that this  high velocity gas  has velocities 
extended beyond those allowed by Galactic rotation.  They designated this feature as Forbidden Velocity Wing 172.8+1.5, which is 
composed of knots, filaments, Sharpless  H\,{\sc ii} regions distributed along a radio continuum loop of size $4^\circ.4 \times 3^\circ.4$. 
They concluded that the H\,{\sc i} gas is well correlated with the radio continuum loop and both of them seem to trace  an expanding 
shell. The expansion velocity and kinetic energy of the shell is estimated as 55 km s$^{-1}$ and $2.5\times10^{50}$ erg, suggesting 
that it could be due to a SN explosion. The authors proposed that the progenitor may have belonged to a stellar association near 
the center of the shell, and this SN explosion triggered the formation of the H\,{\sc ii} regions. 

The  H\,{\sc ii} complex G173+1.5 associated with the Forbidden Velocity Wing features seems to be an 
excellent site to study SN-triggered star formation. A very general description of many star forming regions (SFRs)  in Auriga is 
given in \citet{2008hsf1.book..869R}, and the individual regions have been discussed in several previous studies \citep[e.g.,][]{1983A&A...124....1L, 
1996ApJ...464L.175H,2008MNRAS.384.1675J,2013ApJ...764..172P,2009MNRAS.400..518M,2016MNRAS.459..880M}. 
By mapping the spatial distribution of YSOs around $l\sim$173$^\circ$, $b\sim 1.5^\circ$, 
we can study the star formation in this whole region;  
it is, however, a challenging task to procure uniform optical data over such a large region ($\sim 4^\circ.4 \times 3^\circ.4$).  
The Kiso Wide field Camera \citep[KWFC,][]{2012SPIE.8446E..6LS}, mounted on the Kiso Schmidt telescope operated by the Institute of 
Astronomy of the University of Tokyo, Japan, covers a FOV of $2^\circ.2\times2^\circ.2$ and enables to provide homogeneous 
optical data for a very wide area such as the region of the H\,{\sc ii} complex G173+1.5. Thus by combining $V\&I$ optical photometry 
obtained with this facility with archival data, we can analyze the distribution of the low-mass YSOs and gas/dust around 
$l\sim173^\circ$ in Auriga, aiming to study the star formation in this region. We also attempt to understand the origin and 
evolution of the expanding H\,{\sc i} shell. As a result, we have found two ring-like structures of YSOs/gas/dust distribution extending 
over an area of a few degrees. They are termed as Auriga Bubbles 1 and 2. In this paper, however, we deal mainly with the 
former, and Auriga Bubble 2 will be discussed in a forthcoming paper. 
Section 2 describes the observations, data reduction and completeness of the data.  
The YSO identification technique,  the resulting YSOs sample and their  physical parameters are derived in Section 3. 
In Section 4 we discuss the characteristics and distribution of identified YSOs,  
distribution of gas and dust  as well as  the star formation scenario in the region. We will conclude in Section 5.

\section{Observations and data reduction}

\subsection{Kiso Observations }

\begin{figure*}
\centering
\includegraphics[height=16cm,width=16cm]{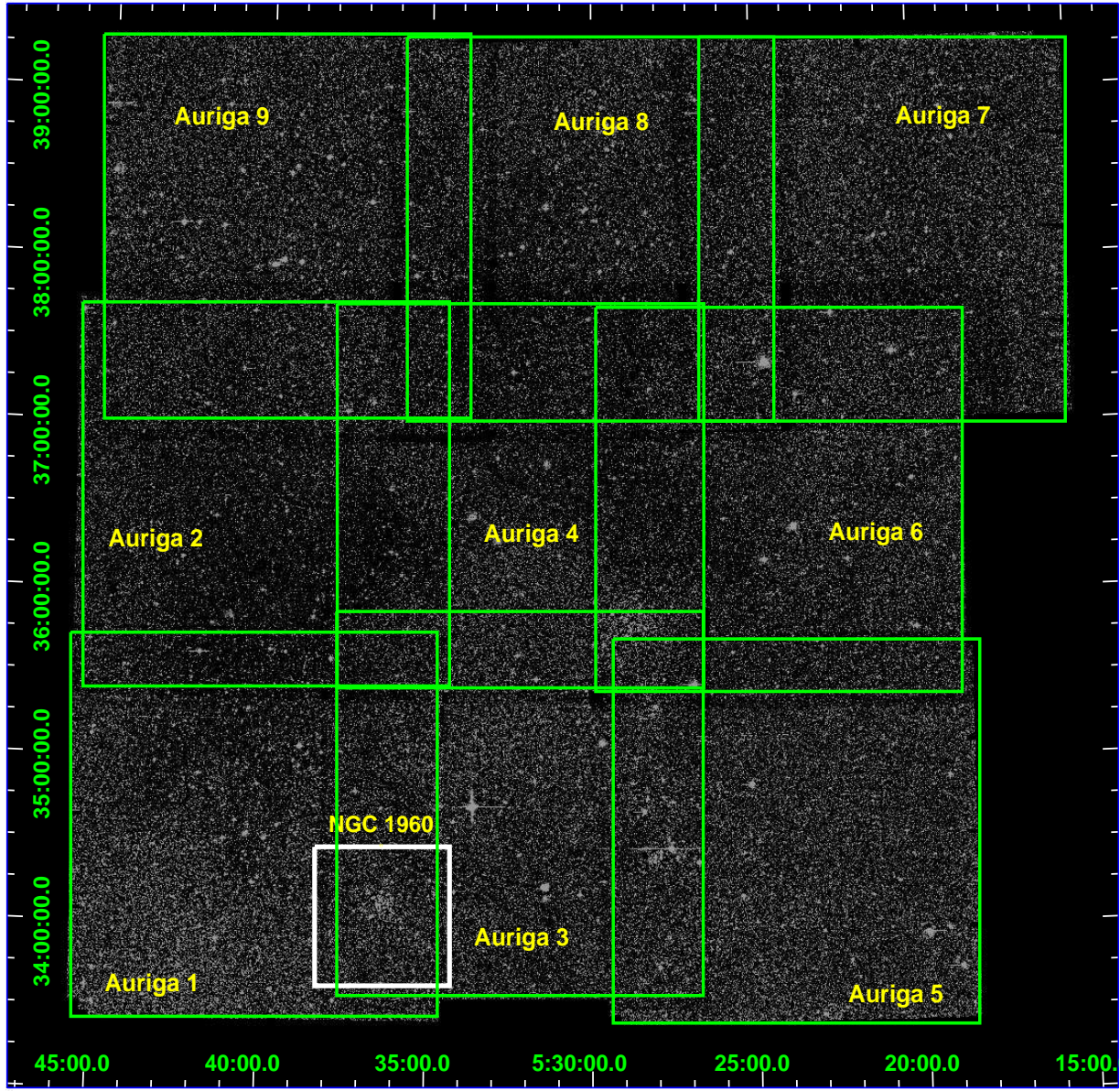}
\caption{\label{spa} The observed region ($\sim 6\times 6$ degree$^2$) of Auriga. 
The nine pointings (green boxes) along with the standard field (NGC 1960: white box) are also shown.
}
\end{figure*}

Optical ($V\&I$) data for nine overlapping regions of Auriga (Auriga 1 - Auriga 9, cf. Fig. \ref{spa}) 
were procured with KWFC \citep[FOV $\sim$2.2$^\circ$ $\times$ 2.2$^\circ$; scale 0.946 arcsec/pixel;][]{2012SPIE.8446E..6LS} on 
the 1.05-m Schmidt telescope at Kiso Observatory, Japan during the  nights of 06, 07 December 2012 and 27 December 2014.  
KWFC, an optical wide-field 64 megapixel imager with 4 SITe and 4 MIT Lincoln Laboratory (MIT/LL) 15$\mu$m 2K$\times$4K 
CCDs is attached to the prime focus of the Kiso Schmidt telescope. The details of the instrument can be found in 
\citet[]{2012SPIE.8446E..6LS} and \citet{2014PASJ...66..114M}. The observations were carried out in the unbinned and slow 
readout mode. In this mode, the readout noise is about 20 and 5-10 electrons for the SITe and MIT/LL CCDs, respectively. A number 
of short and deep exposures were taken. The detailed log of the observations is given in Table \ref{Tlog}. Several bias and dome-flat 
frames were also taken each night.

Initial processing (bias subtraction, flat-fielding and cosmic ray correction) of the data frames was done by using the 
IRAF\footnote{IRAF is distributed by National Optical Astronomy Observatories, USA}
data reduction package. Astrometric calibration and image co-addition were carried out by using the 
SCAMP\footnote{http://www.astromatic.net/software/scamp} and SWarp\footnote{http://www.astromatic.net/software/swarp} software, 
respectively. The total FOV observed in this study is  $\sim6^\circ\times6^\circ$ and is shown in Fig. \ref{spa}. Photometry  of the 
cleaned frames was carried out by using DAOPHOT-II software which includes {\sc find, phot, psf} and {\sc allstar} sub-routines 
\citep{1987PASP...99..191S,1994PASP..106..250S}. The point spread function (PSF) was obtained for each frame by using several 
uncontaminated stars. When brighter stars were saturated on deep exposure frames, their magnitudes have been taken from short 
exposure frames. We used the DAOGROW program for construction of an aperture growth curve required for determining the 
difference between the aperture and PSF magnitudes.

To calibrate the data we used the following procedure. 
We first  calibrated the data for the Auriga 1 field (see Fig.  \ref{spa}) using the photometric data of NGC 1960 taken from 
\citet{2006AJ....132.1669S} which provides photometry in the $U,B,V,R$ and $I$ bands down to a limiting magnitude of 
$V$=20 mag (error$<$0.1 mag) in a FOV of $\sim 1^\circ \times 1^\circ$. This cluster lies in the Auriga 1 field . The area covering 
the observations of NGC 1960 is also shown in Fig. \ref{spa}.

\begin{figure*}
\centering
\includegraphics[height=10cm,width=16cm,angle=0]{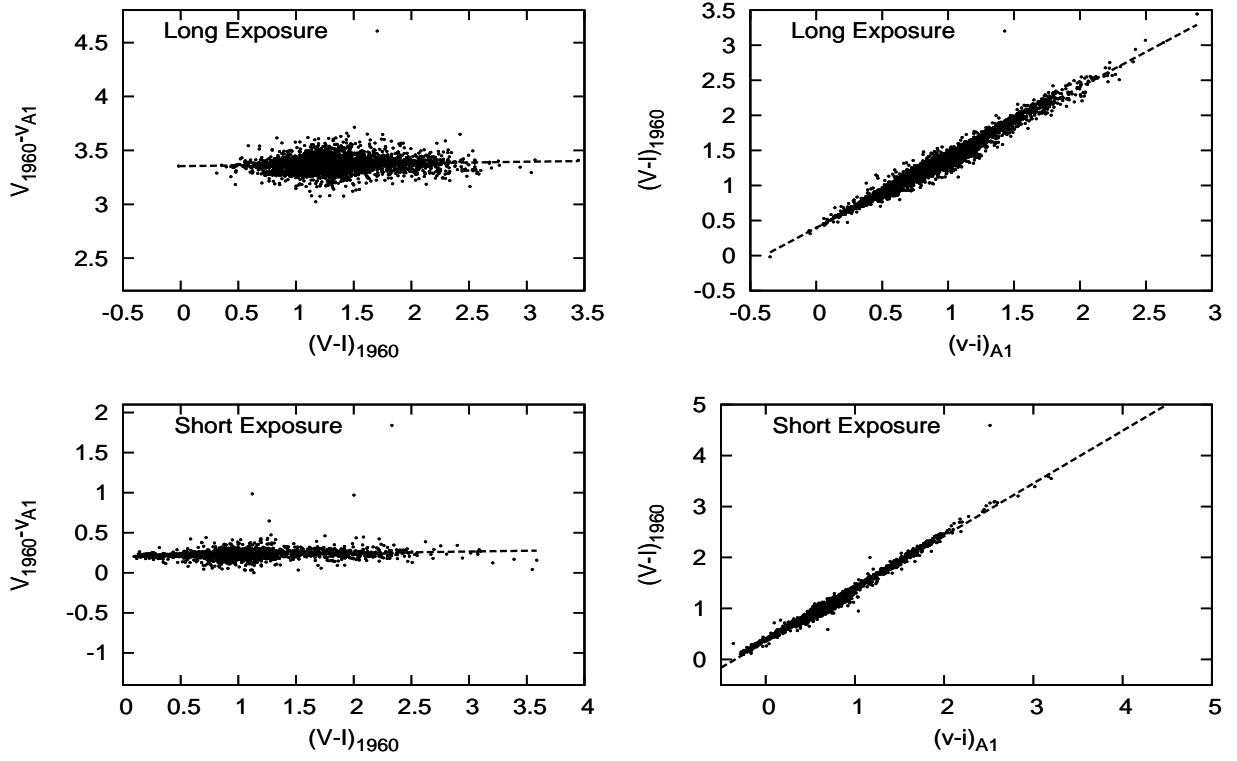}
\caption{\label{calib} Calibration of the Auriga Bubble1 region using the stars in the cluster 
NGC 1960 from \citet{2006AJ....132.1669S} as standard stars for long and short exposures separately. 
The subscripts 1960 and A1 represent the calibrated magnitudes/colours of stars in NGC 1960 
and the present instrumental
magnitudes/colours of stars in Auriga A1, respectively.
 }
\end{figure*}

\begin{figure*}
\centering
\includegraphics[height=16cm,width=10cm,angle=-90]{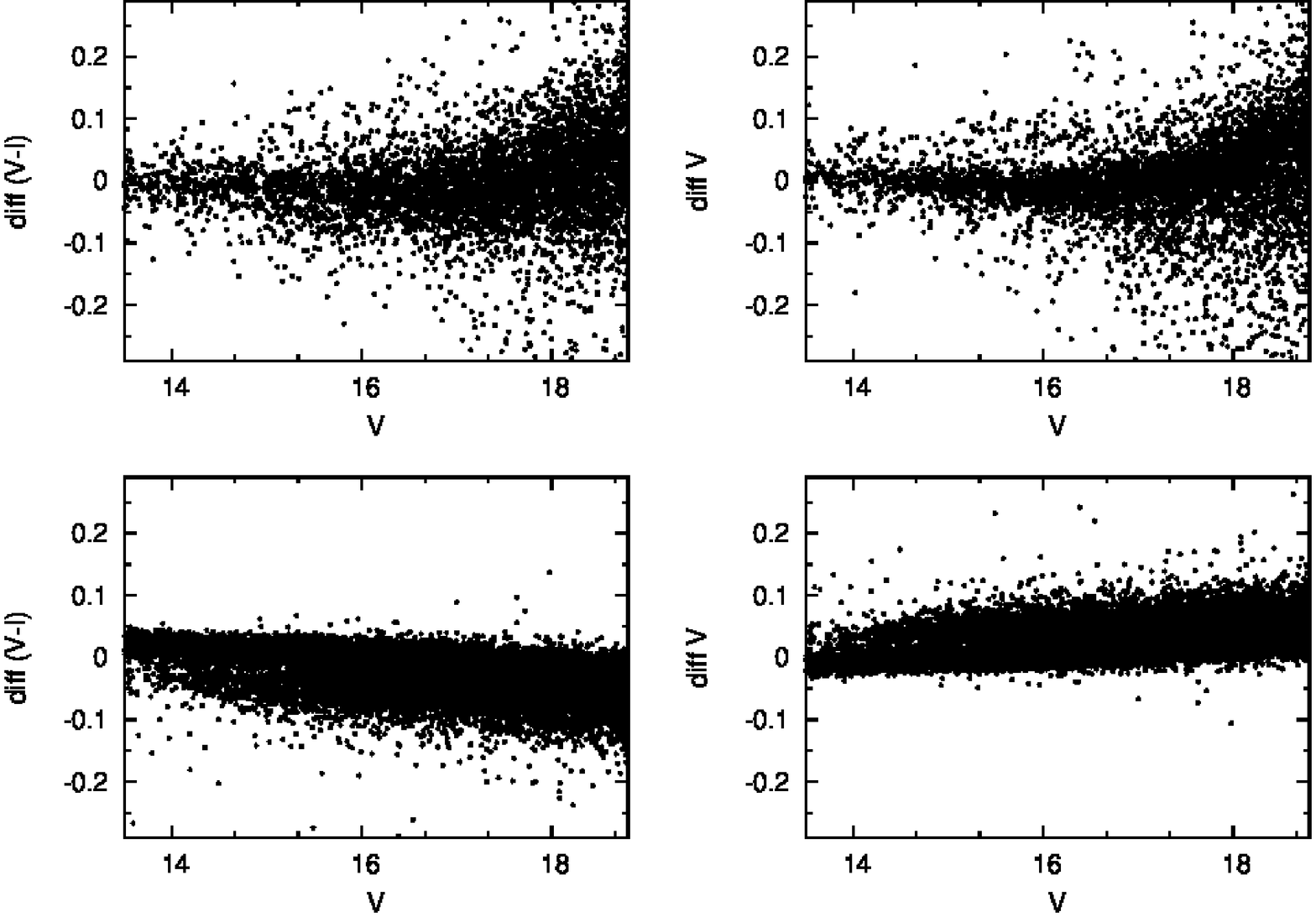}
\includegraphics[height=5cm,width=16cm]{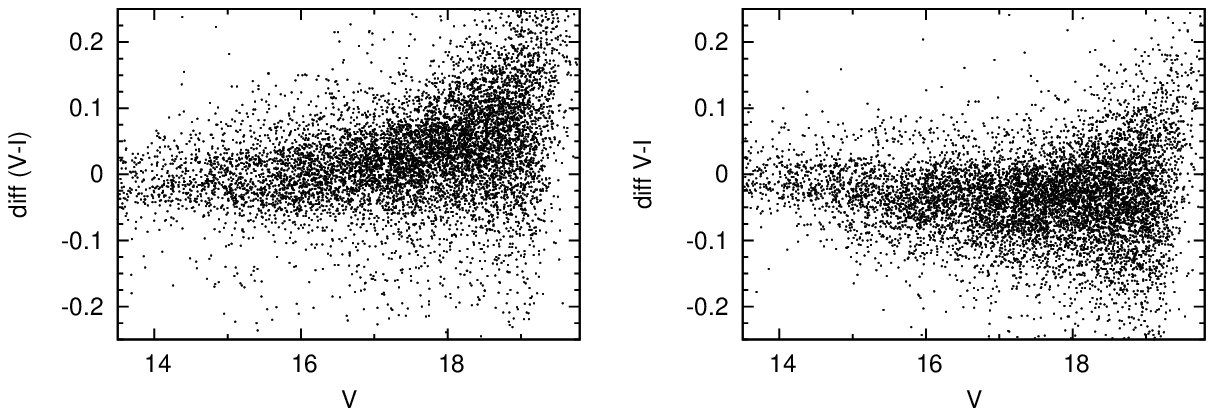}
\caption{\label{resd}  
Differences between the calibrated colours and magnitudes as a function of $V$ magnitude. 
Top panels show the difference between the NGC 1960 cluster data in Auriga 1 field 
(calibrated by using the data by \citet{2006AJ....132.1669S}) and the NGC 1960 data, 
whereas middle panels plot the difference between the Auriga 1 field data calibrated by using NGC 1960 
and those calibrated by using the common stars in Auriga 1 and Auriga 3, cf. Table 3). 
The lower panel compares the photometry of common stars (calibrated by using the NGC 1960 data) of 
Auriga 1 and Auriga 3 (cf. Table \ref{cmpT1}).
}
\end{figure*}

After calibrating the Auriga 1 field, stars lying in the common overlapping area of Auriga 1  and Auriga 2   
were taken as the secondary standards to calibrate  the  Auriga 2 field. The common stars of  Auriga 2 and Auriga 4 were taken 
as the secondary standards to calibrate  the  Auriga 4 field  and so on.  
To translate the instrumental magnitudes to the standard magnitudes the following calibration equations, derived by using a 
least-square linear regression, were used: 

\begin{equation}
	V - v = M1\times(V-I) + C1
\end{equation}
\begin{equation}
	(V-I) = M2\times(v-i) + C2,
\end{equation}

\noindent where $V$ and $(V - I)$ are the standard magnitudes and colours from \citet{2006AJ....132.1669S}, respectively;  $v$, 
and $(v - i)$ are the instrumental magnitudes and colours, respectively;  C1, C2  and  M1, M2  are the zero-point constants and 
colour coefficients, respectively.  
Fig. \ref{calib} plots, as an example, the $(V-I)$ vs. $(V-v)$ and $(v-i)$ vs. $(V-I)$ 
diagram  for the stars in NGC 1960. 
Fig.  \ref{resd} (top panels) plot the difference between the NGC 1960 data from
\citet[][]{2006AJ....132.1669S} and the calibrated $V$ magnitudes 
and $(V-I)$ colours of the NGC 1960 region obtained by using the equations 1 and 2.
The standard deviation in $\Delta V$ and $\Delta (V - I)$, in the magnitude range $V \sim$ 13.5 - 19.0 mag, is 0.09 and 0.08 mag, 
respectively. The values of all the zero-point constants and  colour coefficients for various fields (Auriga 1 - Auriga 9) are given in 
Table \ref{coeff}.

\begin{figure*}
\centering
\includegraphics[width=0.45\textwidth,angle=0]{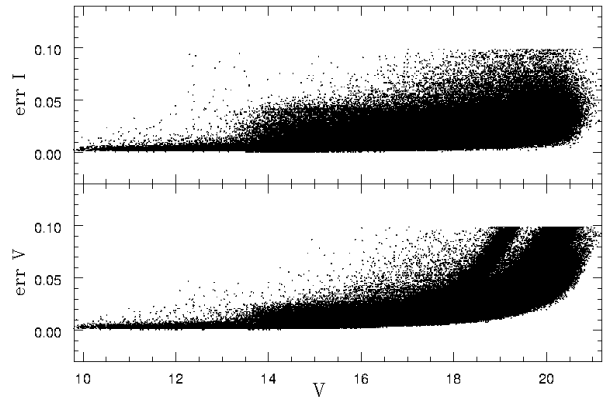}
\caption{\label{err} DAOPHOT errors as a function of $V$ magnitude.
The data for $V<$13.5 and $V>$13.5 are taken from short and long exposures, respectively.} 
\end{figure*}

The Auriga 3 field  was calibrated by using the
common stars in Auriga 5 and Auriga 3 as secondary standards as described earlier in this section. 
We further used the common stars lying in Auriga 1 and Auriga 3 to get another set of the 
transformed standard magnitudes for Auriga 1 stars. Fig. \ref{resd} (middle panel) and Table \ref{cmpT} 
shows the residuals  $\Delta$, between the two sets of calibrated 
data of Auriga 1 in $V$ magnitudes and $(V - I)$ colours. 
The standard deviation in $\Delta V$ and $\Delta (V - I)$, in the magnitude range $V \sim$ 13.5 - 19.0 mag 
is $\sim$0.02 and $\sim$0.03 mag, respectively. 
Although there is some trend in lower panel, but the scatter is small. 
The comparison of these two transformed standard data indicates a fair agreement.  
Here we would like to point out that although the uncertainty due to the calibration shown in Fig. \ref {calib}
is not negligible, the uncertainty in optical data due to calibration are not crucial for the scientific results of the present study as 
they can be safely assumed to be negligible compared to those associated with 
the analysis of the Spectral Energy Distribution (SED) aimed at 
the derivation of stellar and disk parameters (cf. Section 3.3.1).
To ensure the quality of the calibration of the data we further calibrated the 
Auriga 3 field using the NGC 1960 data from \citet[][]{2006AJ....132.1669S} 
and compared the calibrated phtometric data of common stars in the FOVs of Auriga 1 and Auriga 3.  
The comparison is shown in the lower panel of Fig. \ref{resd}. 
Table \ref{cmpT1} shows residuals between the two photometric calibrations. The standard deviation 
in $\Delta V$ and $\Delta (V - I)$, in the magnitude range $V\sim$13.5 - 19.0 mag is 
$\sim$0.07 and $\sim$0.08 mag, respectively. The comparison of these two independently calibrated data  
indicates a fair agreement and ensures the reliability of the calibration.  
The typical DAOPHOT errors in magnitude  as a function of the corresponding magnitudes are shown in Fig. \ref{err}.
It can be seen that the errors become large ($\sim$0.1 mag) for stars  $V \sim 20$ mag. 
More than 0.47 million stars have errors less than 0.1 mag 
in $V$ and $I$ bands and these stars have been used for analysis in the ensuing sections.

\subsection{Archival data: 2MASS and $WISE$  data \label{obs-spit}}

To investigate the star formation  in the Auriga Bubble region we also used near/mid 
infra-red (NIR/MIR) data from the Two Micron All Sky Survey \citep[2MASS,][]{2003yCat.2246....0C} and the Wide-field Infrared Survey 
Explorer survey \citep[$WISE$,][]{2010AJ....140.1868W} archived data in addition to the optical $VI$ data from the Kiso Schmidt 
telescope.

$WISE$ is a 40-cm telescope in low-Earth orbit that surveyed the whole  sky in four mid-IR bands at 3.4, 4.6, 12, and 22 $\mu$m 
(bands $W1$, $W2$, $W3$, and $W4$) with nominal angular resolutions  of 6$^\prime$$^\prime$.1, 6$^\prime$$^\prime$.4, 
6$^\prime$$^\prime$.5, and 12$^\prime$$^\prime$.0 in the respective bands \citep{2010AJ....140.1868W}.  In this paper, we make 
use of the AllWISE  source catalog, which provides accurate positions, apparent motion measurements, four-band fluxes and flux 
variability statistics for over 747 million objects. The online explanatory supplement\footnote{http://wise2.ipac.caltech.edu/docs/release/allwise/expsup/} 
and Marsh \& Jarrett (2012) describe the $WISE$ source detection method in detail. The AllWISE catalog is searchable via 
NASA/IPAC Infrared Science Archive (IRSA). It also provides information on 2MASS counterparts.

We have also used  the 2MASS Point Source Catalog \citep[PSC,][]{2003yCat.2246....0C} for near-IR (NIR) 
$JHK$$_s$ photometry of all the sources (including those sources which do not have $WISE$ photometry) in the Auriga Bubble 
region. This catalog is reported to be 99$\%$ complete down to the limiting magnitudes of 15.8, 15.1 and 14.3 in the $J$, $H$ and
$K_s$ band, respectively \footnote{http://tdc-www.harvard.edu/catalogs/tmpsc.html}.

\subsection{Completeness of the data}

To have an unbiased study of star formation in the region, it is vital to know the completeness of 
 the data in terms of magnitudes/ masses for the sample YSOs identified. The photometric data may be incomplete due to various 
reasons, e.g., nebulosities,  crowding of stars, detection limit etc \cite[][]{1991A&A...250..324S,2008AJ....135.1934S}.

\begin{figure*}
\centering
\includegraphics[width=0.45\textwidth]{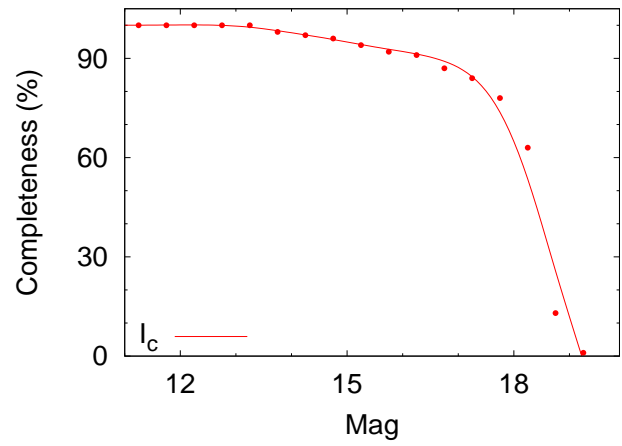}
\caption{\label{fcft} 
Completeness levels for $I_c$  band as a function of magnitude
derived from the artificial star experiments (see Section 3.2 for details).
}
\end{figure*}

 To check the completeness factor  for the optical data we used the ADDSTAR routine of DAOPHOT II. This method has been used 
by various authors \citep[cf.][and references therein]{2007MNRAS.380.1141S,2008AJ....135.1934S}. Briefly, the method consists of 
randomly adding artificial stars of known magnitudes and positions into the original frame. The frames are reduced by using the 
same procedure used for the original frame. The ratio of the number of stars recovered to those added in each magnitude interval 
gives the  completeness factor as a function of magnitude. The luminosity distribution of artificial stars is chosen in such a way that 
more stars are inserted into the fainter magnitude bins. In all about 15\% of the total stars are added so that the crowding 
 characteristics of the original frame do not change significantly \citep[see][]{1991A&A...250..324S}. 
 We found that the present optical data for $I$ band are complete at $\sim90\%$ level for 17.75 magnitude (cf. Fig. \ref{fcft}).

 The estimate of the completeness of the present YSO sample is rather difficult as it is limited by several factors. For example, the 
bright  nebulosity in the $WISE$ bands significantly limits the point source detection. The YSO identification on the basis of 
2MASS - $WISE$ colours may be limited by the sensitivity of the 2MASS survey and saturation of $W3$ and $W4$ images.  
Variable reddening and stellar crowding characteristics across the region could also affect the local completeness limit. 
The completeness of the YSO selection using the NIR data  is also  
dictated by the amount of IR excess, the evolutionary status of disks etc.  
In the present study the larger $WISE$ PSF may also hamper the detection of YSOs in the crowded region. 
 All these effects are difficult to quantify.

\section{Results}

\subsection{YSO identification \label{idf}}

In the present study the 2MASS data along with the $WISE$ data have been used to identify and classify the YSOs associated 
with the Auriga Bubbles 1 and 2 using the following classification schemes.
YSOs are generally classified as Class\,{\sc 0}, Class\,{\sc i},  Class\,{\sc ii} or Class\,{\sc iii} 
sources on the basis of the infrared slopes of their spectral energy distributions \citep{1994ASPC...65..197B,2006AJ....131.1574L}.
YSOs of the early stages (i.e., Class\,{\sc 0} and {\sc i}) are usually deeply buried inside the molecular clouds, hence detection of 
them at optical wavelengths is  difficult. The most prominent feature of these YSOs is accreting circumstellar disks and 
envelopes. The radiation from the central YSO is  absorbed by the circumstellar material and re-emitted in NIR/MIR. 
Therefore, Class\,{\sc i} (and Class\,{\sc ii} as well) sources can  be probed through their IR excess (compared to normal 
stellar photospheres).

\subsubsection{$WISE$ classification}

The four wave-bands of $WISE$ are useful to detect mid-IR emission from cold circumstellar disk/envelope material in YSOs.  This 
makes $WISE$ all-sky survey as a  readily available tool to identify and classify YSOs, in a similar way to what can be done with 
$Spitzer$ \citep[][and others]{2004ApJS..154..363A,2008ApJ...674..336G,2009ApJS..184...18G} but over the entire sky. We use 
the AllWISE Source Catalog to search for YSOs adopting the criteria given by \citet{2014ApJ...791..131K}. We refer 
Figure 3 of \citet{2014ApJ...791..131K} to summarize the entire scheme. This method includes the selection of candidate 
contaminants (AGN, AGB stars and star forming galaxies). Out of 40788 $WISE$ source meeting photometric quality flags, 
22238 sources were excluded from the data file as contaminants on the basis of selection criteria of  \citet{2014ApJ...791..131K}.
In the present study we followed and applied the photometric quality criteria for different $WISE$ bands as 
given in \citet{2014ApJ...791..131K}.

\paragraph{$WISE$ three-band classification}

\begin{figure*}
\centering
\includegraphics[height=8cm,width=8cm]{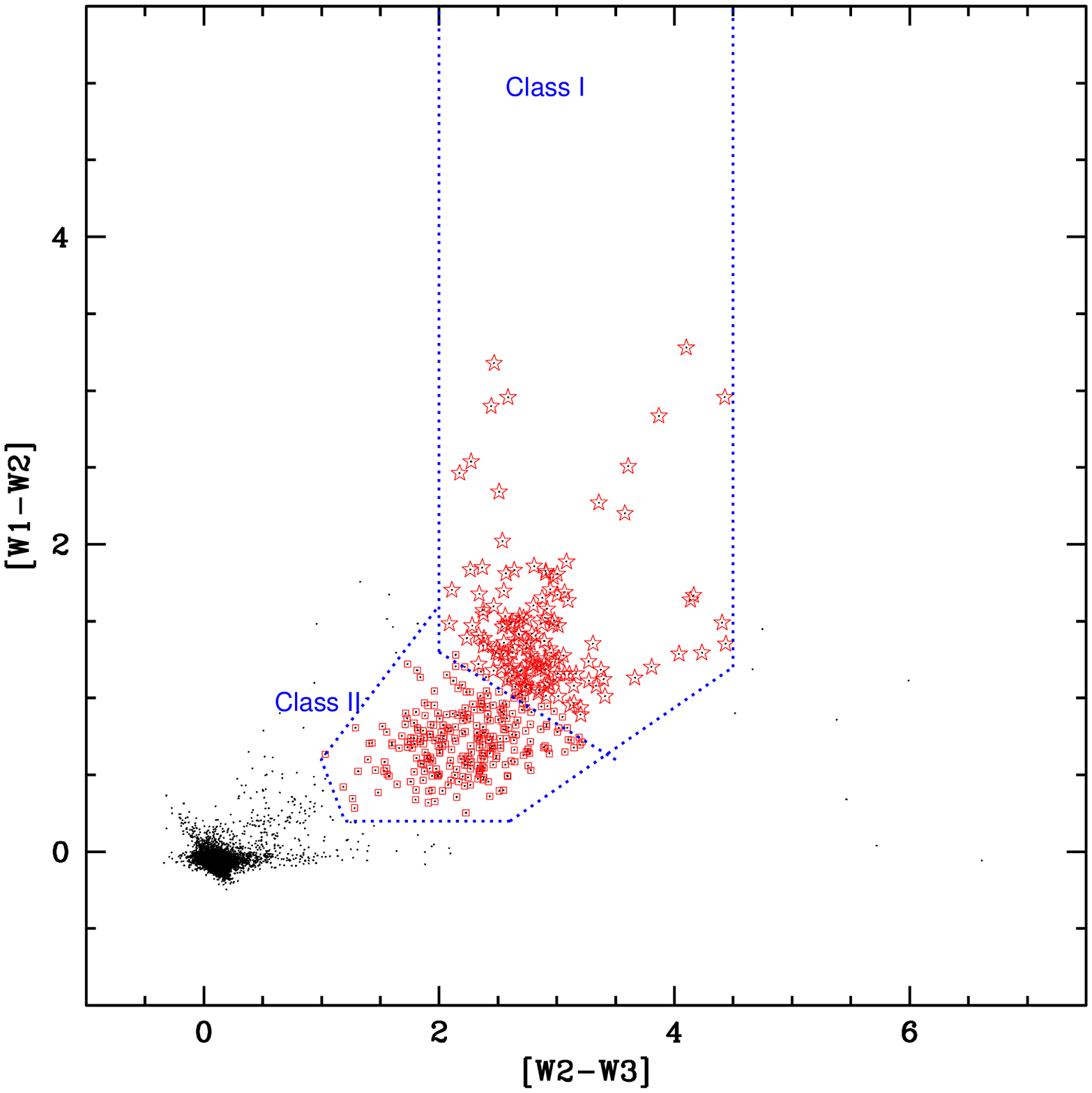}
\includegraphics[height=8cm,width=8cm]{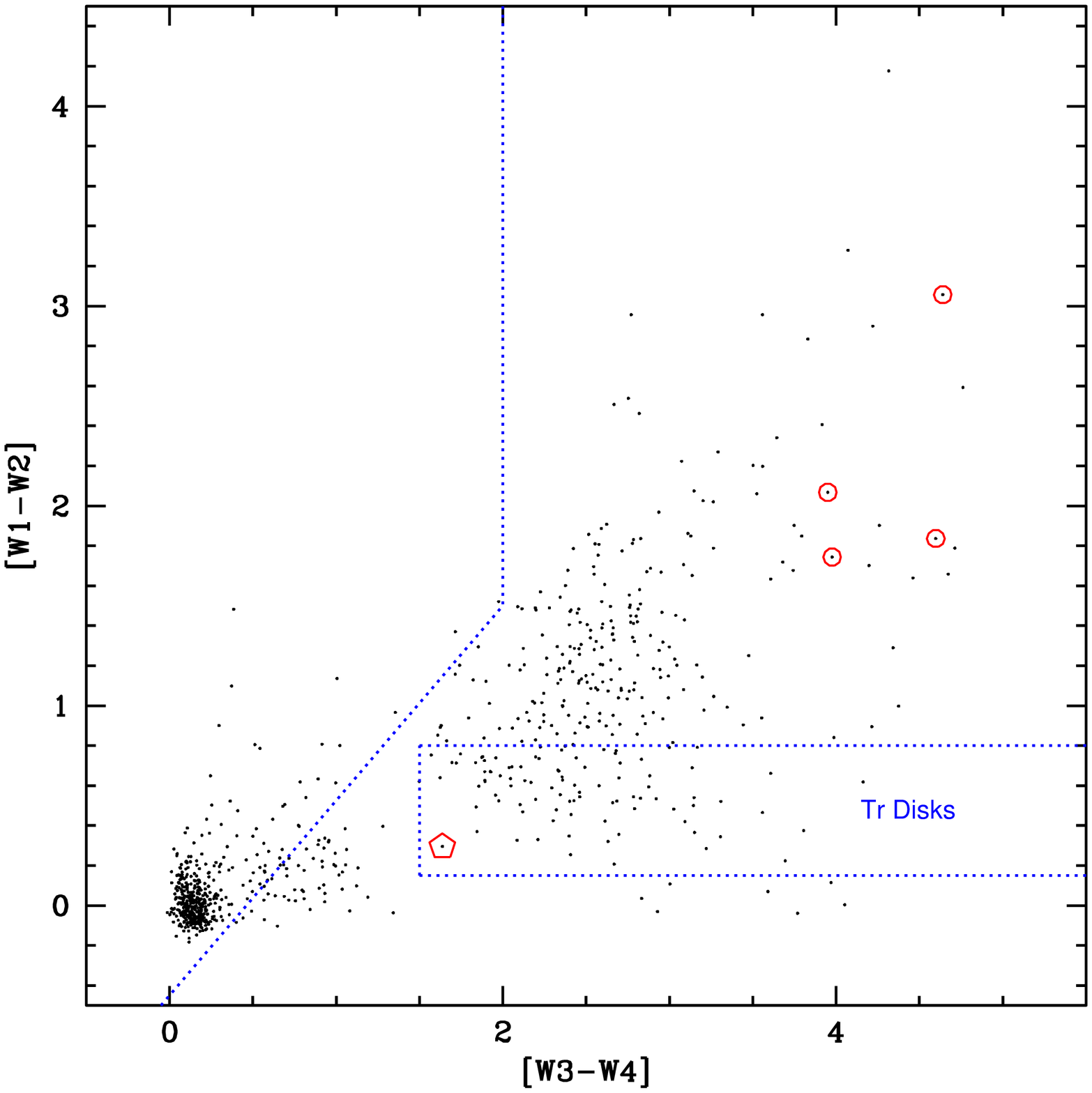}
\includegraphics[height=8cm,width=8cm]{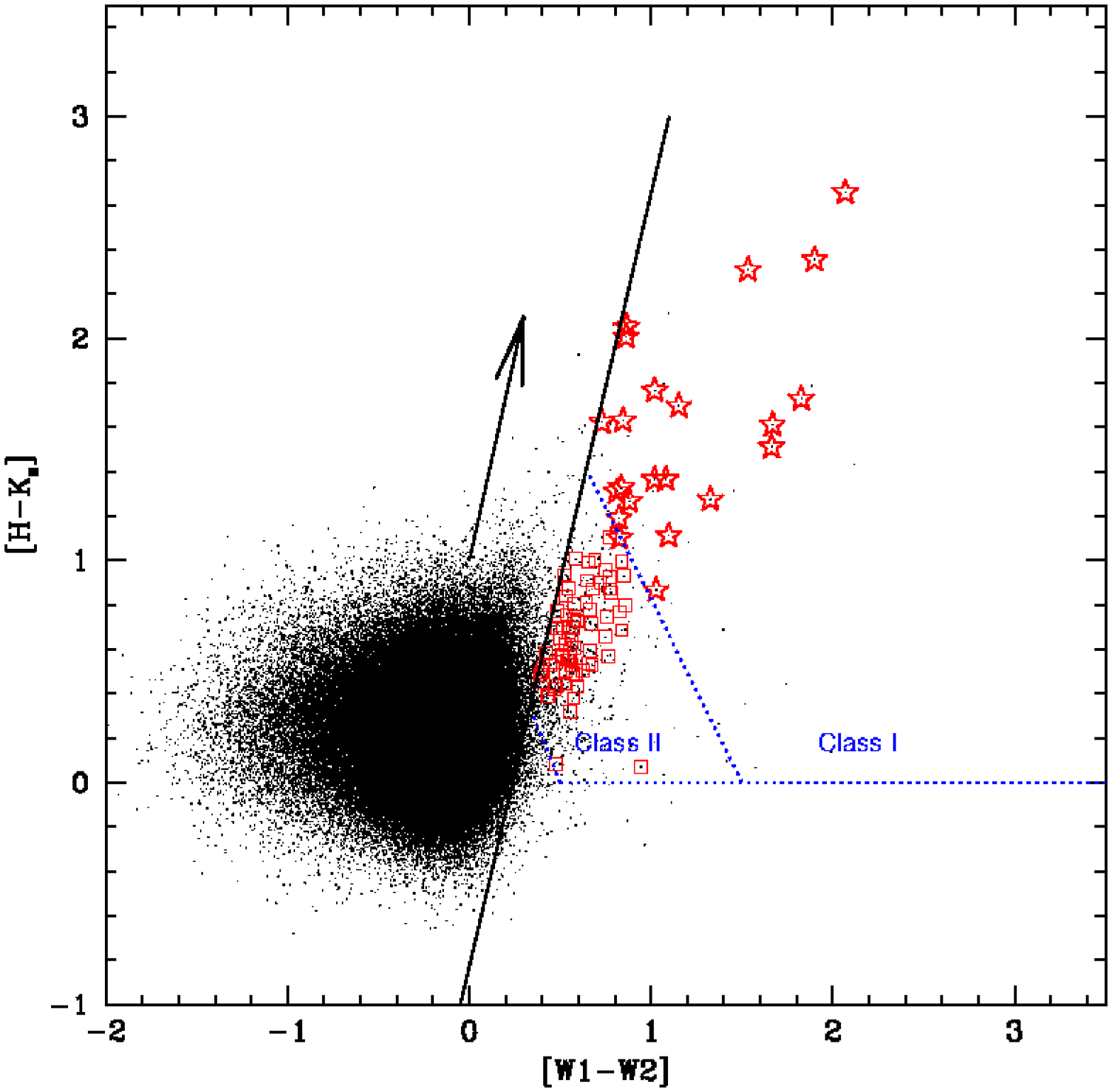}
\includegraphics[height=8cm,width=8cm]{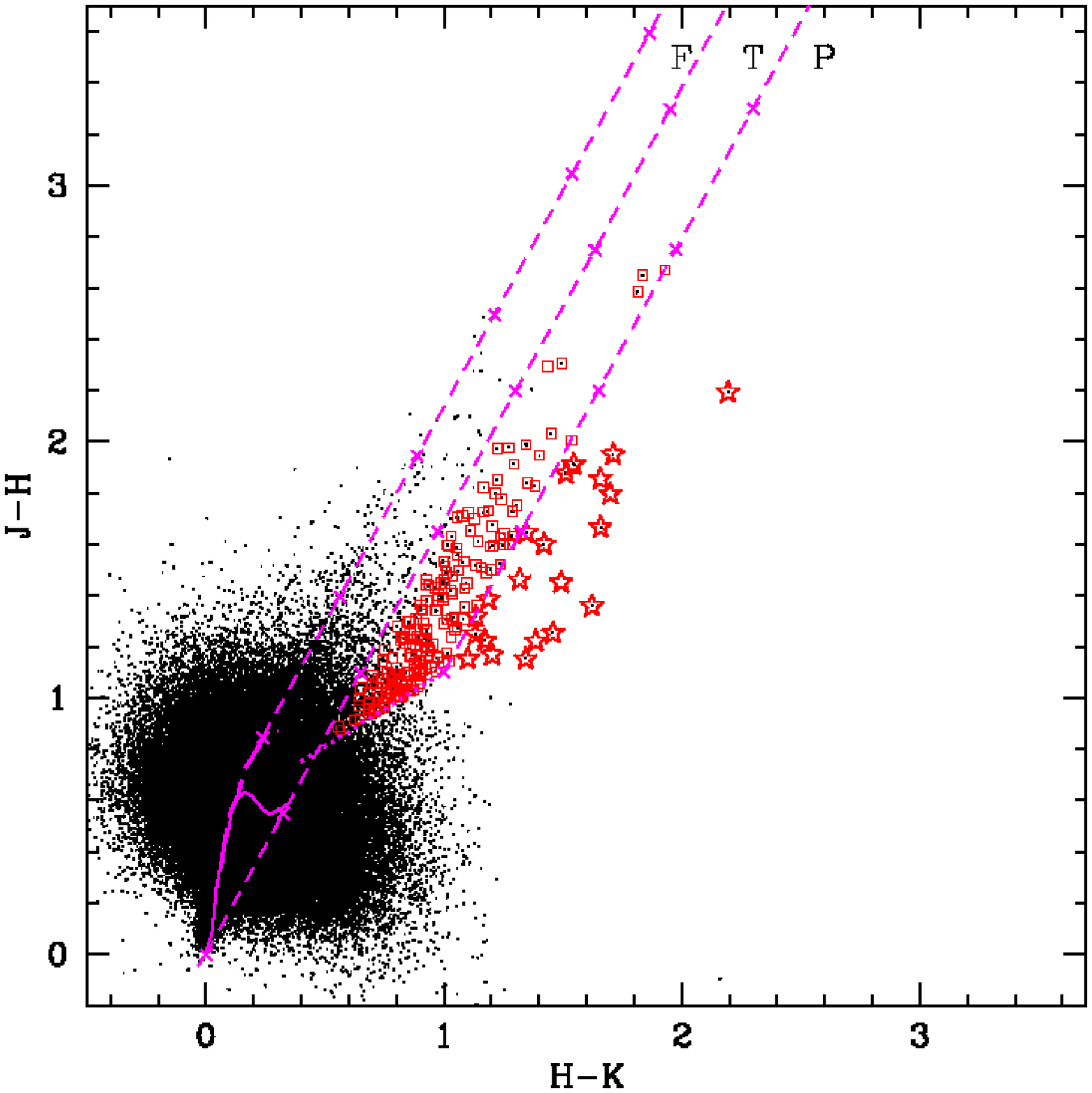}
\caption{\label{wise} WISE/2MASS TCDs of all the sources within the studied region (black dots). The
YSOs classified as Class\,{\sc i}, Class\,{\sc ii}, and transition disk stars
 on the basis of colour criteria by \citet{2014ApJ...791..131K} and \citet{2004ApJ...608..797O}
are shown with stars, squares, and pentagon symbols, respectively.
Dotted lines show the YSO Class divisions.
An arrow in bottom-left panel shows the extinction vector of $A_{KS}$ = 2 mag. 
In the same panel, the solid black line parallel to the extinction vector represents  the highest 
range of extinctions tabulated in Table 10 of \citet{2014ApJ...791..131K}.
The continuous and thick  magenta dashed curves in the bottom right-hand panel 
represent the intrinsic main sequence (MS) and giant branches \citep{1988PASP..100.1134B},
respectively. The dotted line indicates the loci of unreddened CTTSs \citep{1997AJ....114..288M}.
The parallel magenta dashed lines in the same panel 
are the reddening lines drawn from the tip (spectral type M4) of the
giant branch (left reddening line), from the base (spectral type A0) of the MS branch (middle reddening
line) and from the tip of the intrinsic CTTS line (right reddening line).
The crosses on the reddening lines show an increment of $A_V$ = 5 mag. }
\end{figure*}

We applied this classification scheme  to all the sources that were detected in three  $WISE$ 
bands (namely $W1$, $W2$ and $W3$) and satisfying the photometric quality criteria   as 
given in \citet{2014ApJ...791..131K}. The YSOs were selected on the basis of their  $(W1 - W2)$ and  $(W2 - W3)$ colours 
according to  the criteria  given by \citep{2014ApJ...791..131K}, which are based on the colours of known YSOs in Taurus, 
extragalactic sources, and Galactic contaminants.  This approach efficiently separates out IR excess 
contaminants such as star forming galaxies, broad-line active galactic nuclei, 
unresolved shock emission knots, objects that suffer from polycyclic aromatic hydrocarbon (PAH) 
emissions etc. \citep[see e.g., figure 2 of][]{2016ApJ...827...96F}. Fig. \ref{wise} (top left-hand panel)  shows the $(W2 - W3)$ versus
$(W1 - W2)$ $WISE$ two-colour diagram (TCD) for all the sources in the region, 
where candidate  YSOs classified as Class\,{\sc i} and Class\,{\sc ii} are shown by red stars and  red squares, respectively. 
However, it is worthwhile to mention that there could be a overlap between 
Class\,{\sc i} and Class\,{\sc ii} objects as can be seen  in figure 5 of  \citet{2014ApJ...791..131K}. 
However, this is not expected to affect  our discussion on bubbles as we are using both 
Class\,{\sc i} and Class\,{\sc ii} YSOs having age $\leq$ 1 Myr to delineate the structure of the bubbles
(cf. Section 4.3, Fig. \ref{gal}).

\paragraph{$WISE$ four-band classification}

$W4$ photometry has been used to identify  candidates transition disk objects 
and also to retrieve candidate protostars which might have been classified as 
AGN candidate on the basis of the $WISE$ three band classification scheme (cf. Sec. 3.1.1.1).
Fig. \ref{wise} (top right-hand panel) shows the $(W3 - W4)$ versus  $(W1 - W2)$ TCD
for all the sources in the region, where probable YSOs classified as transition disk sources and protostars
are shown by red pentagons and  red circles, respectively.
The classification discussed in this section yields one and four transition disk source and protostars, respectively.

\paragraph{2MASS and $WISE$ classification }

Since the studied region has highly variable nebulosity, many sources in the 
region may not be detected at longer wavelengths due to the saturation of detectors, 
hence the selection of  YSOs on the basis of only $WISE$ bands photometry may not be complete.  
Therefore, we use 2MASS $H$, $K_s$ data along with $W1$ and $W2$ band $WISE$ data 
as some sources might have not been 
detected in the $W3$ band, but have possibility of detection 
in the 2MASS $H$ and $K$ band \citep{2014ApJ...791..131K}. Fig. \ref{wise} (bottom left-hand panel) shows the $(W1 - W2)$ 
versus  $(H - K)$ TCD for all the sources in the region, where probable YSOs classified as Class\,{\sc i} and Class\,{\sc ii} are shown 
by red stars and  red squares, respectively.
All the YSOs identified above have been checked against the possibility of being AGB stars as discussed by 
\citet{2014ApJ...791..131K}. Finally on the basis of the $WISE$ data we classified 154 and 331 sources as  probable Class\,{\sc i} 
and Class\,{\sc ii} YSOs. 

\subsubsection{2MASS classification}

The  $(J - H)/(H - K)$ NIR TCD is also a useful tool to identify pre-main sequence (PMS) objects. It is possible that some of the 
candidate YSOs might have not been detected on the basis of $WISE$ data, hence we have also used 2MASS $JHK$ three band 
data to identify additional YSOs.  Fig.~\ref{wise} (bottom right-hand panel) displays the NIR TCD for all  the stars which have not 
been detected in the WISE survey or do not meet the photometric quality criterion (cf. Section 3.1.1) in the region studied. 
The 2MASS magnitudes and colours have been converted into the California Institute of Technology (CIT) 
system\footnote{http://www.astro.caltech.edu/$\sim$jmc/2mass/v3/transformations/}. The solid and long dashed lines in Fig. 
\ref{wise} (bottom left-hand panel) represent unreddened main sequence (MS) and giant branch loci \citep{1988PASP..100.1134B}, 
respectively.  The dotted line indicates the intrinsic loci of CTTSs \citep{1997AJ....114..288M}. 
The parallel dashed lines are the reddening vectors drawn from the tip (spectral type M4) of the giant branch (`left reddening line'), 
from the base (spectral type A0) of the MS branch (`middle reddening line') and from the tip of the intrinsic CTTSs  line (`right 
reddening line'). The extinction ratios $A_J /A_V$ = 0.265, $A_H /A_V$ = 0.155 and $A_K /A_V$ = 0.090 have been adopted from 
\citet{1981ApJ...249..481C}. The sources lying in the `F' region could be either field stars (MS stars, giants), Class\,{\sc iii} or 
Class\,{\sc ii} sources with small NIR excesses. The sources lying in the `T' region are considered to be mostly classical T-Tauri 
stars (CTTSs, i.e., Class\,{\sc ii} objects). The sources lying in the `P' region - redward of the right reddening line -  are most likely 
Class\,{\sc i} objects \citep[protostellar-like objects;][]{2004ApJ...616.1042O}. In  this scheme we  consider only those sources as 
YSOs that lie at a location above the intrinsic loci of CTTSs with margins larger than the errors in their colours.
This classification criterion yields 21 and 204 probable Class\,{\sc i} and Class\,{\sc ii} YSOs, respectively. 

\subsection{YSO sample}

On the basis of the criteria discussed above  we have compiled a catalog of 710 YSOs in an area of  
 $\sim$6$\times$6 degree$^2$, divided into  175 Class\,{\sc i} and 535 Class\,{\sc ii} YSOs.  
A portion of the catalog is shown in Table \ref{data1_yso}, which lists the positions of YSOs, their magnitudes at various 
 bands, and classification.  The complete  catalog is available in an electronic form only. 
The optical magnitudes of the nearest optical counterparts for 176  YSOs which have been found within a match radius of 2 
arcsec are also given in Table \ref{data1_yso}.
Here, is it worthwhile to mention that no multiple identification between the optical and IR counterparts of YSOs
are found within 2 arcsec.

\begin{figure*}
\centering
\includegraphics[width=0.75\textwidth]{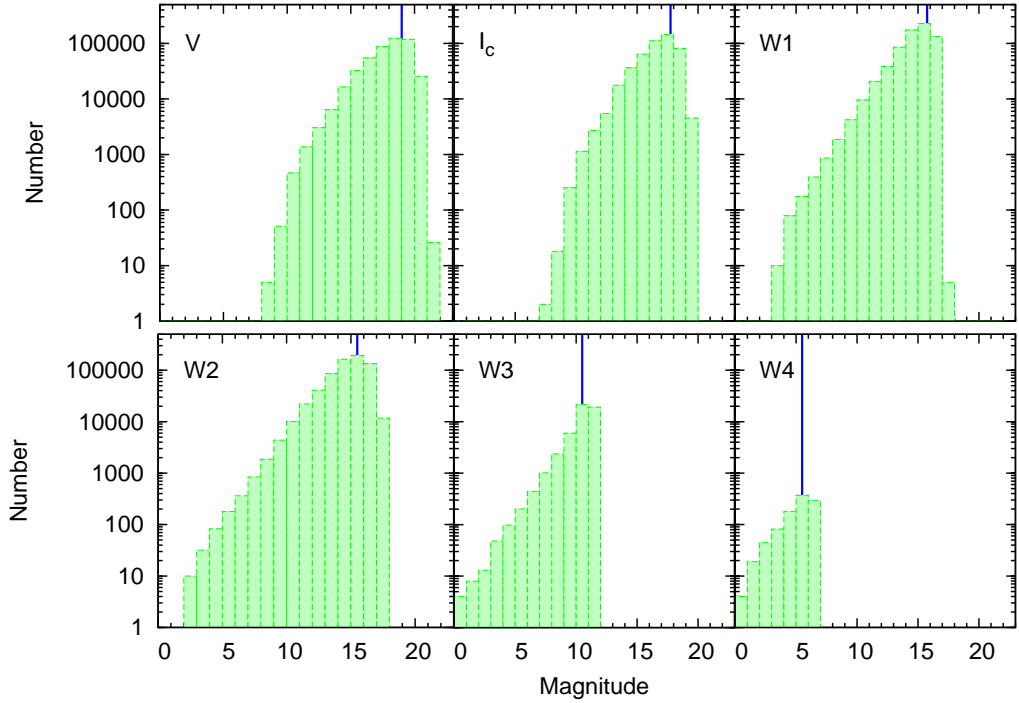}
\caption{\label{cft-hist} Histograms of the source numbers for different bands showing the limiting
magnitude and completeness limit for each band. The vertical lines indicate the
adopted completeness limit.
}
\end{figure*}

 As assumed in various previous studies, the peak of the observed luminosity function can be 
 considered as the 90\% completeness limit of the data \citep[cf.][]{2003PASP..115..965E, 
 2013ApJ...778...96W, 2013MNRAS.432.3445J, 2016ApJ...822...49J, 2016AJ....151..126S}. 
 We constructed the  luminosity function for each band (see Fig. \ref{cft-hist}). The resultant completeness limits of the optical and 
$WISE$ data are given in Table \ref{cftT}. 
The completeness limit in the $I$ band obtained here agrees well with that obtained in Sec 2.3.
As mentioned in Sec 2.2, we assume that the 2MASS $JHK$ data have completeness 
 of $\sim 90\%$ at the limiting magnitudes of 15.8, 15.1 and 14.3 for the $J, H$ and $K$ bands respectively.

\subsubsection{Unidentified Class\,{\sc iii} (diskless) sources}

In the present study we have not attempted to identify diskless YSOs. The Class\,{\sc iii} sources  
may possibly have a different spatial distribution in the Auriga region as compared to the Class\,{\sc i} and \,{\sc ii} 
YSOs and inclusion of these sources may have impact on the parameters described in the ensuing sections. 
It is worthwhile to mention that disk fraction estimates in young clusters having 
age $\leq 3$ Myr varies from 60-70\% (M $\lesssim$ 2 M$_\odot$) and 35 - 40\% 
(M$>$2 M$_\odot$) \citep{2015A&A...576A..52R}, whereas the disk half-life estimates 
varies from 1.3 Myr to 3.5 Myr \citep[][and references therein]{2018MNRAS.477.5191R}.  
Since the missing YSO mass due to the incompleteness of our YSO search criteria 
as well as to unidentified Class\,{\sc iii} sources may play a role 
in the analysis to be carried out in the ensuing sections, we will assume 
that 50\% of the total YSO population is missed in the present YSO sample.

 \subsection{Physical properties of the identified YSOs}

 \subsubsection{Spectral energy distribution  }

  The YSOs can also be characterized  from their SED.  The SED fitting provides evolutionary stages 
and physical parameters such as mass, age, disk mass, disk accretion rate and photospheric temperature of YSOs and hence is 
an ideal tool to study their evolutionary status.  
We constructed the SEDs of the YSOs using the grid models and Python version of 
SED fitting tools\footnote{https://sedfitter.readthedocs.io/en/stable/} of 
\citet{2006ApJS..167..256R,2007ApJS..169..328R}\footnote{WISE fluxes were acquired from Dr. Thomas Robitaille through private communication}.
The models were computed by using a Monte-Carlo based 20000 2-D radiation 
transfer calculations from \citet{2003ApJ...591.1049W,2003ApJ...598.1079W,2004ApJ...617.1177W} and by adopting several 
combinations of a central star, a disk, an infalling envelope, and a bipolar cavity in a reasonably large parameter space and with 
10 viewing angles (inclinations). 

\begin{figure*}
\centering\includegraphics[height=6cm,width=7cm,angle=0]{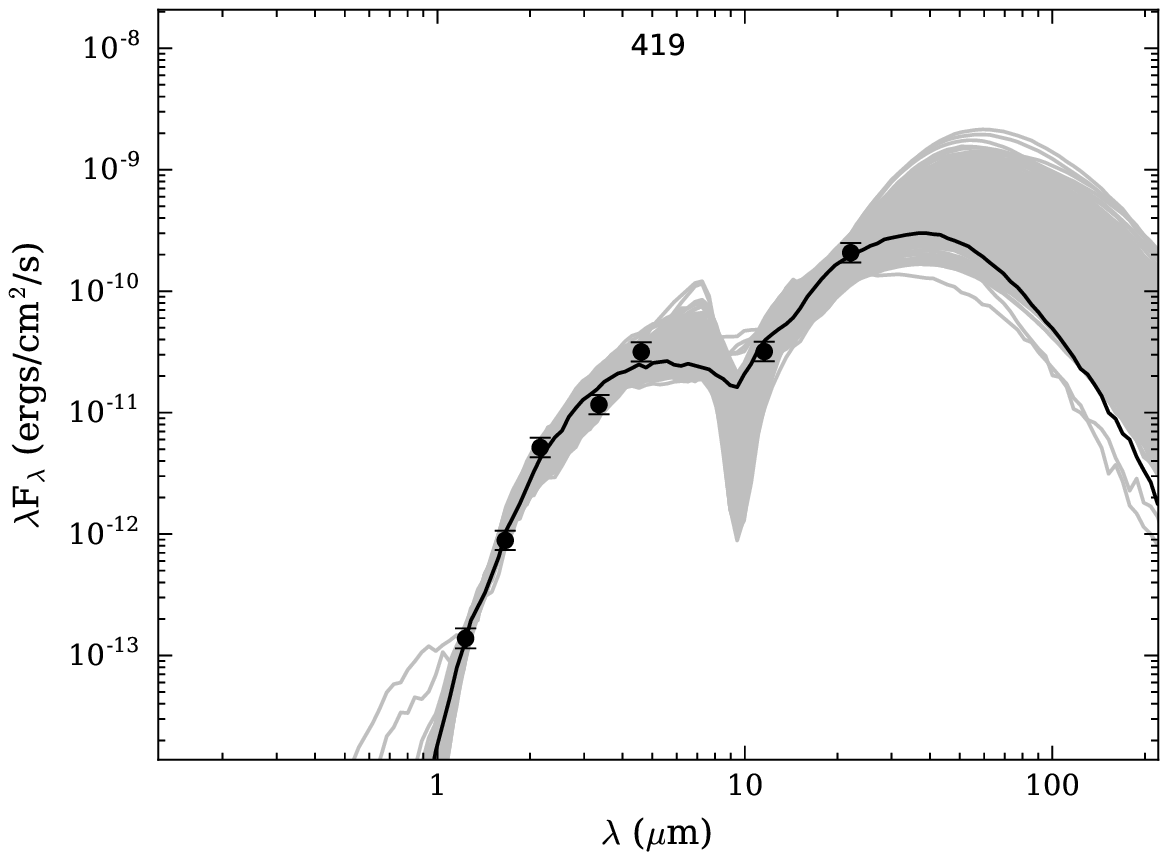}
\centering\includegraphics[height=6cm,width=7cm,angle=0]{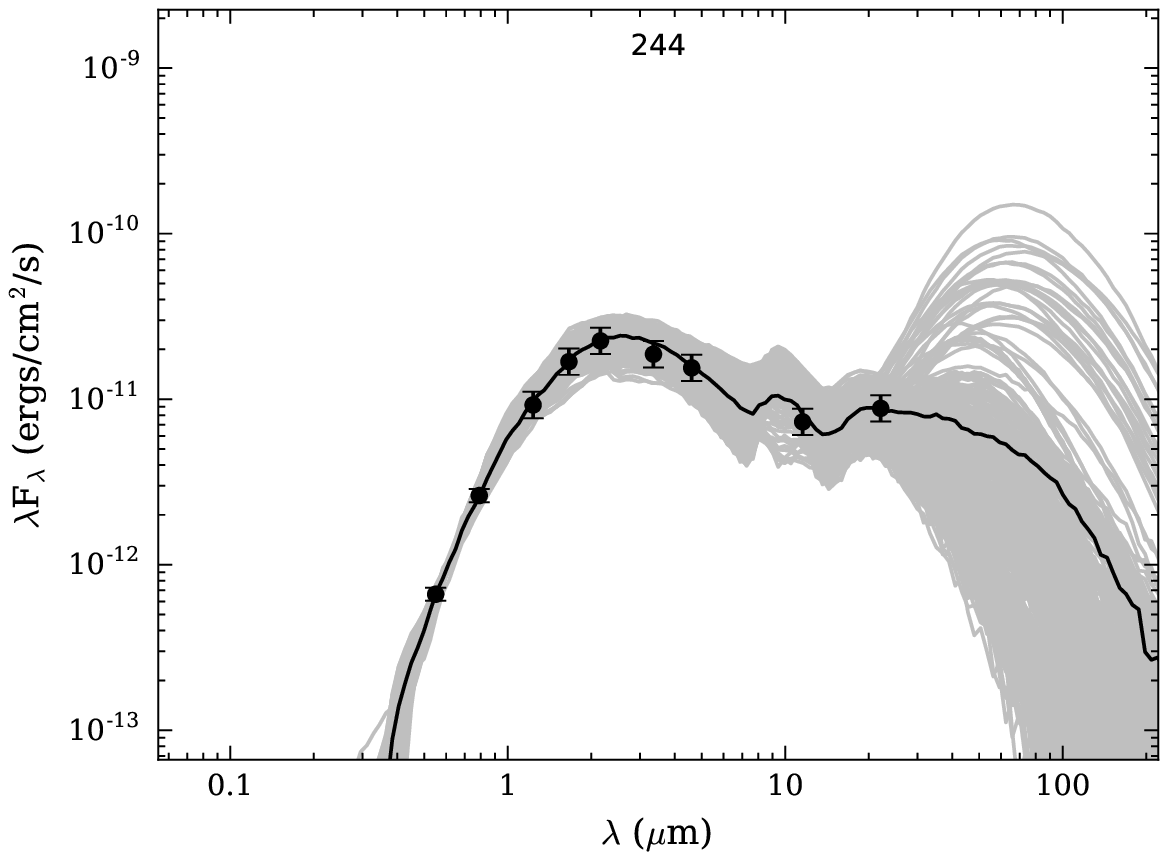}
\caption{\label{sed} Sample SEDs for Class\,{\sc i} (left-hand panel) and Class\,{\sc ii} (right-hand panel)  sources
, respectively, created by the SED fitting tools of \citet{2007ApJS..169..328R}.
The black curve shows the best fit and the gray curves show the subsequent well fits 
satisfying our selection criteria (see text for details).
The filled circles with error bars denote the input flux values.}
\end{figure*}

  The SEDs were constructed by using the multiwavelength data (i.e. optical to MIR) with the condition that a minimum of 5 data 
points should be available. While fitting the models to the data  we assumed the extinction and the distance as free parameters. 
Considering the errors associated with the distance estimates available in the literature, 
the range in distance estimate was assumed to vary between 2.0 kpc to 2.4 kpc. 
Since the extinction  in the region is variable \citep[cf.][]{2008MNRAS.384.1675J, 2013ApJ...764..172P}, we used a range for  
$A_V$ of 1.6 to 30 mag. We further set photometric uncertainties of 10\% for optical and 20\% for both NIR and MIR.  These values 
are adopted instead of the formal errors in the catalog in order to fit without any possible bias in  underestimating the flux 
uncertainties. In Fig. \ref{sed}, we show example SEDs of Class\,{\sc i} and Class\,{\sc ii} sources, where the solid black curves 
represent the best-fit and the gray curves are the subsequent well-fits satisfying our requirements for good fit discussed below. 

Since the SED models are highly degenerate, the best-fit model is unlikely to give an unique solution and the estimated physical 
parameters of the YSOs tabulated in Table \ref{data3_yso} are the weighted mean with standard deviation of the physical 
parameters obtained from the models that satisfy $\chi^2 - \chi^2_{min} \leq 2 N_{data}$, where $\chi^2_{min}$ is the goodness-of-fit 
parameter for the best-fit model and $N_{data}$ is the number of input data points. 

  \section{Discussion}

  \subsection{Physical conditions in the Auriga region}

  In this Section, we present a brief description of the identified YSOs, ionized gas and molecular clouds, along with their correlation 
with each other. The region contains five Sharpless H\,{\sc ii} regions Sh2-231 to Sh2-235. In addition it is reported that there are 
about 14 embedded SFRs having ages $\sim$3-5 Myrs \citep[cf.][and references therein]{2012AJ....143...75K}. It  has also been 
proposed that the formation of these young objects could have been triggered by a older generation of stars
\citep[][and references therein]{2012AJ....143...75K}.

  \subsubsection{Characteristics of the identified YSOs in the Auriga region}

\begin{figure*}
\centering
\includegraphics[height=4cm,width=5.7cm]{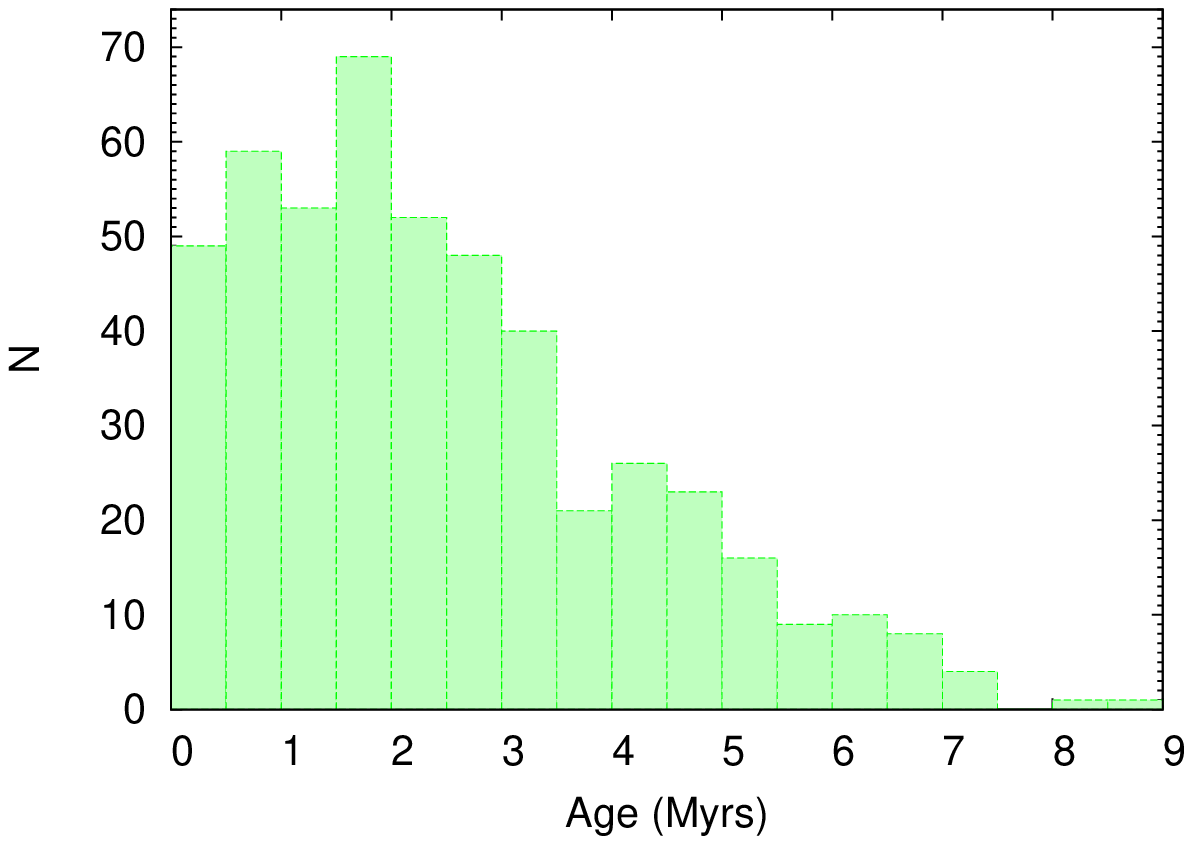}
\includegraphics[height=4cm,width=5.7cm]{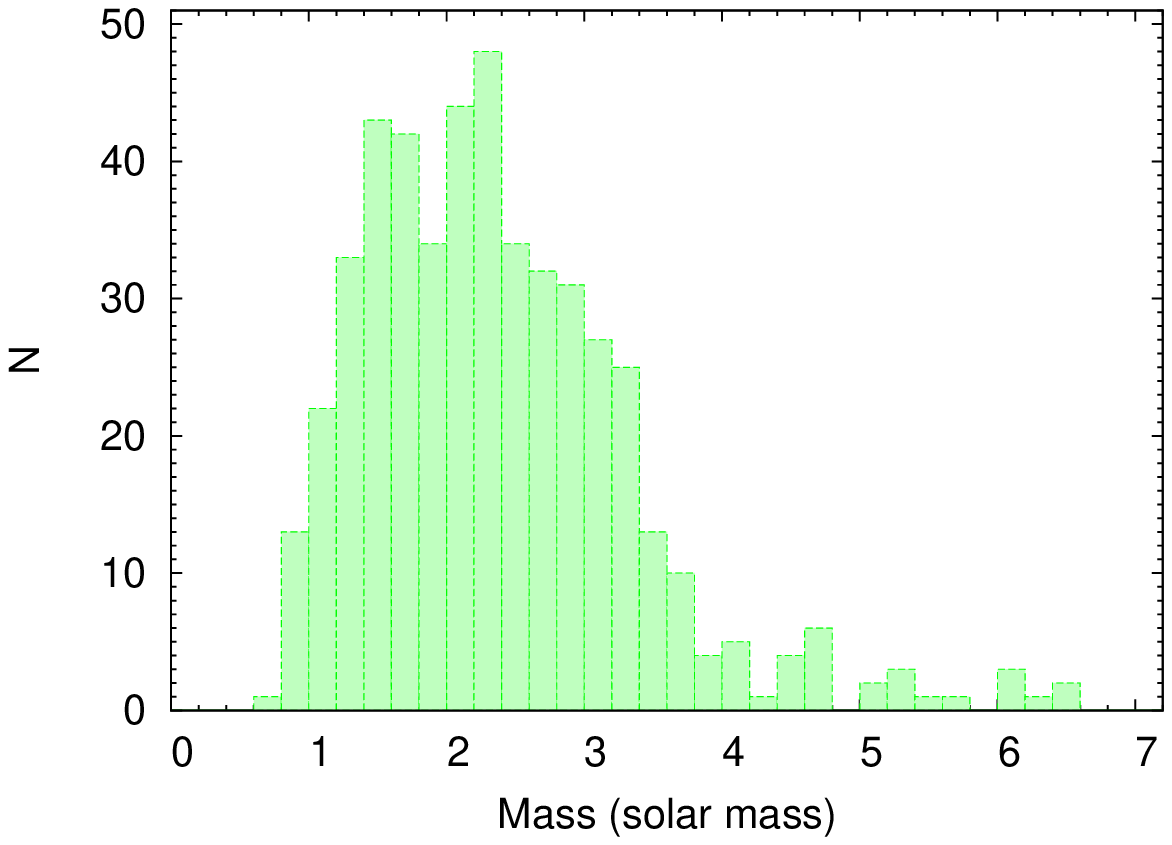}
\includegraphics[height=4cm,width=5.7cm]{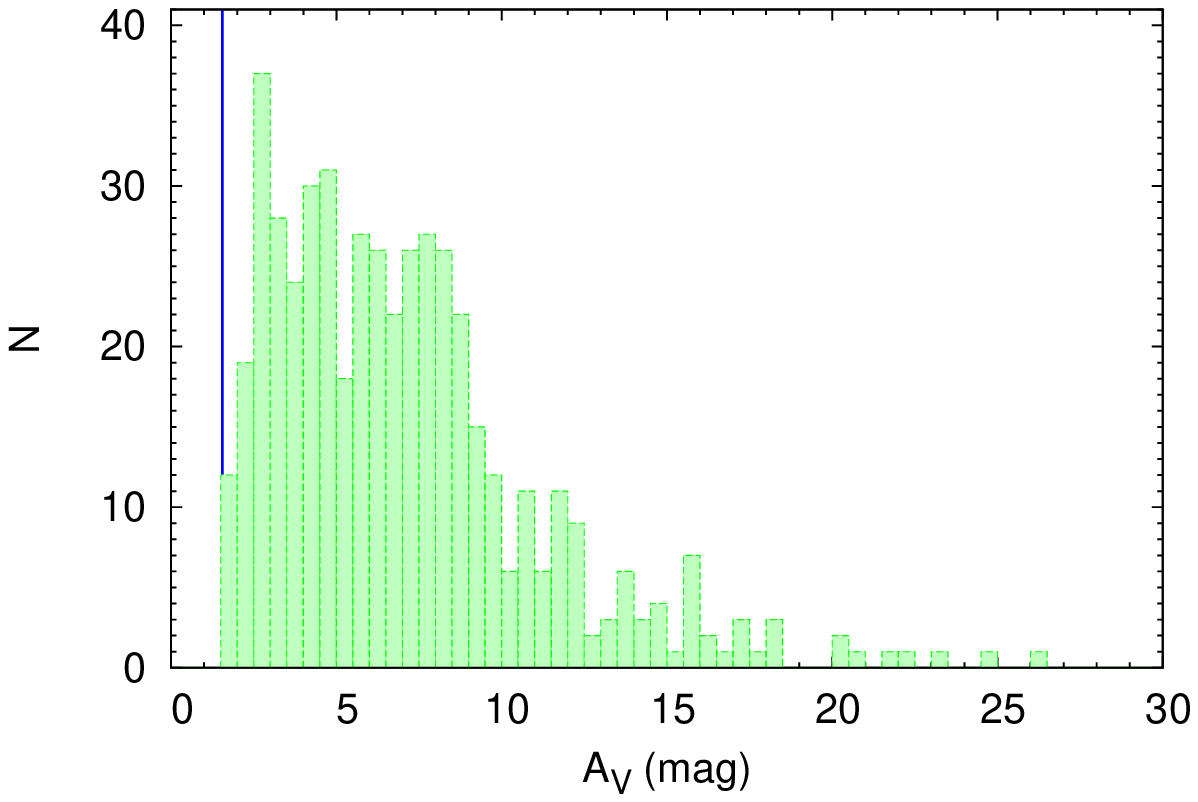}
\caption{\label{histogram} Histograms showing the distribution of the ages (left panel),
masses (middle panel) and extinction values `$A_V$'  (right panel), respectively,  of the YSOs in the Auriga Bubble region.
A vertical line in the right panel indicates the foreground reddening value (i.e. $A_V$ = 1.55 mag).
}
\end{figure*}

 We have identified 710 YSOs in the Auriga Bubble region (cf. Table \ref {data1_yso}) 
and derived the physical parameters of  489 YSOs from the SED fitting analysis (cf. Table \ref {data3_yso}).
These parameters were used in further analysis. Histograms of the age and mass of these YSOs are shown in Fig.~\ref{histogram}. 
The distribution of the ages estimated on the basis of SEDs indicates that $\sim$76\% (370/489) of the sources have 
  ages $\le $ 3.5 Myr. The masses of the YSOs are found to range between 0.75 to 9 M$_\odot$, and a majority ($\sim$86\%) of 
them are in the range of 1.0 to 3.5 M$_\odot$. The $A_V$ distribution shows a long tail indicating its large spread from 
$A_V$=1 - 27 mag, which is consistent with the nebulous nature of this region. The average age, mass and extinction ($A_V$) for 
this sample of YSOs are $2.5\pm1.7$ Myr, $2.4\pm1.1$ M$_\odot$ and $7.1\pm4.1$ mag, respectively.

  The evolutionary classes of the identified 710 YSOs given in the Table \ref {data1_yso} reveal that $\sim$25\% (175 out of 710) 
sources are Class\,{\sc i} YSOs.  The comparatively high percentage of Class\,{\sc i} YSOs indicates the youth of this region.  

  \subsubsection{The distribution of gas, dust and YSOs}  

Details on the environment and distribution of YSOs can be used to probe  the star formation scenario in the region. In 
Fig. \ref{Fiso} (top left-hand panel), the  H\,{\sc i} contours by \citet{1990A&AS...85..691F} (black contours) 
and the $^{12}$CO contour map (cyan contours) from \citet{2001ApJ...547..792D} along with the distribution of the YSOs are 
overlaid on the $WISE$ 12 $\mu$m image. The $WISE$ 12 $\mu$m image covers the prominent PAH features at 
11.3 $\mu$m, which is indicative of star formation activity \citep[see e.g.][]{2004ApJ...613..986P}. This figure reveals that ionized 
as well as molecular gas distribution is well correlated with that of YSOs.
The distribution of ionized gas, molecular gas and YSOs indicates a ring-like structure spread 
over an area of a few degrees.
The distribution of YSOs  also reveals that a majority of Class\,{\sc i} sources belong generally 
to this ring-like structure, whereas the  comparatively older population, i.e. Class\,{\sc ii}  
objects, are rather randomly distributed throughout the region. As stated already we term this structure as Auriga Bubble1. 
Furthermore there is  another,  very well  defined distribution of Class\,{\sc i}  and Class\,{\sc ii} 
objects towards    the north-west of the H\,{\sc ii}    complex G173+1.5, forming another bubble feature, which we call as  Auriga  
Bubble2. Its nature will be discussed in an ensuing study.

  \subsubsection{Extinction and YSOs surface density maps}

  To quantify the  extinction in the region and to characterize the structure of the molecular 
gas associated with various SFRs in the Auriga Bubble,  
we derived $A_K$ extinction maps using the $(H - K)$ colours of field  stars \citep[cf.][]{2011ApJ...739...84G}. 
To produce the extinction map we excluded the candidate YSOs (cf. Section 3.2) and probable 
contaminating sources (AGNs, AGB stars and star forming galaxies) using the procedure by \citet{2014ApJ...791..131K} from the 
sample stars. Similar approaches were used by other studies also \citep[e.g.,][and references 
therein]{2008ApJ...675..491A,2009ApJS..184...18G,2013MNRAS.432.3445J,2016ApJ...822...49J,2016AJ....151..126S}.
Mean values of $A_K$  were derived by using the nearest neighbor (NN) method as 
described in detail by \citet{2005ApJ...632..397G} and \citet{2009ApJS..184...18G}. 
Briefly, the mean value of $(H - K)$ colours of five nearest stars at each position in a grid of 30 arcsec 
was calculated for the entire Auriga region ($\sim6^\circ\times6^\circ$). 
The sources deviating above 3$\sigma$ were excluded to calculate the final mean colours of each grid. 
The reddening law $A_K$ =  1.82 $\times$ ($(H - K)_{obs}- (H - K)_{int}$) by \citet{2007ApJ...663.1069F}  was used to convert the 
$(H -K)$ colours into $A_K$, where $(H - K)_{int}$ = 0.2 was assumed as an average intrinsic colours of the field stars 
\citep[see.][]{2008ApJ...675..491A, 2009ApJS..184...18G}. To eliminate the foreground contribution in generating the extinction map 
we used only those stars which have $A_K >$ 0.15$\times$D, where D is the distance in kpc \citep{2005ApJ...619..931I}. 
The extinction map is sensitive down to $A_K\sim$2.8 mag (=$A_V\sim$30 mag). However, the derived $A_K$ values are to be 
considered as lower limits, because the sources in the region with higher extinction might have not been detected in the present 
sample. The extinction map, smoothened to a resolution of 18 arcmin, is shown in blue colour in Fig. \ref{Fiso} (top right-hand panel). 
It is interesting to note that the extinction map resembles the general distribution of the molecular gas as outlined by the $^{12}$CO 
emission map shown in Fig. \ref{Fiso} (top left-hand panel). It shows a concentration of molecular clouds towards Sh2 - 231 - 235. 
A comparison of the stellar density distribution and the morphology of the molecular material can provide a clue to the history of star 
formation in the region. The surface density maps of YSOs were generated by using the NN method as described by  
\citet{2005ApJ...632..397G}. We used the radial distance that contains 5 nearest YSOs to compute the local surface density in a 
grid size of 30 arcsec. The grid size identical to that of the extinction map was used to compare the stellar density and the gas 
column density. The density distribution of YSOs is shown in red colour in Fig. \ref{Fiso} (top right-hand panel). The distribution of 
the YSOs and molecular material shows a nice correspondence. \citet{2011ApJ...739...84G} also found similar trends in eight nearby 
molecular clouds.  \citet{2008ApJ...674..336G} have shown that the  sources in each of the Class\,{\sc i} and Class\,{\sc ii} 
evolutionary stages have very different spatial distributions relative to that of the dense gas in their natal cloud. We have also 
compared the extinction map with the positions of YSOs of different evolutionary stages and found that the Class\,{\sc i} sources are 
located towards the places with higher extinction. These properties agree well with previous findings 
in several star-forming regions such as W5 \citep{2008ApJ...688.1142K,2012A&A...546A..74D}, 
i.e., younger Class\,{\sc i} sources are more clustered and closely associated with the densest molecular clouds in which they were 
born presumably, while the Class\,{\sc ii} sources are scattered probably by drifting away from their birthplaces.

  \subsection{Clustered population  in the Auriga Bubble}

  \subsubsection{Extraction of sub-clusters and the distribution of scattered YSO  population}

\begin{figure*}
\centering
{
\hspace{-0.3 cm}
\includegraphics[width=0.515\textwidth]{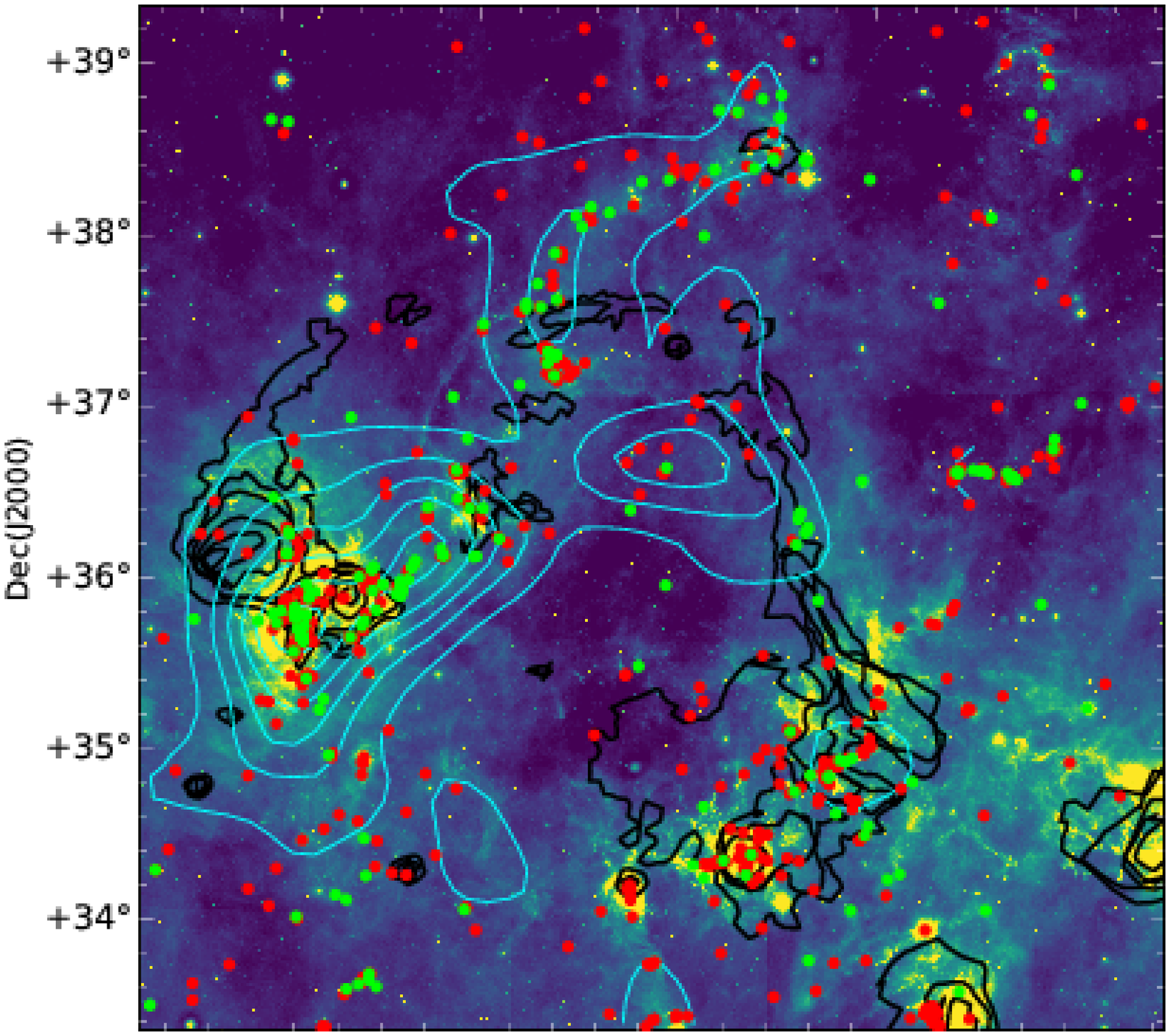}
\includegraphics[height=8.0cm,width=8cm]{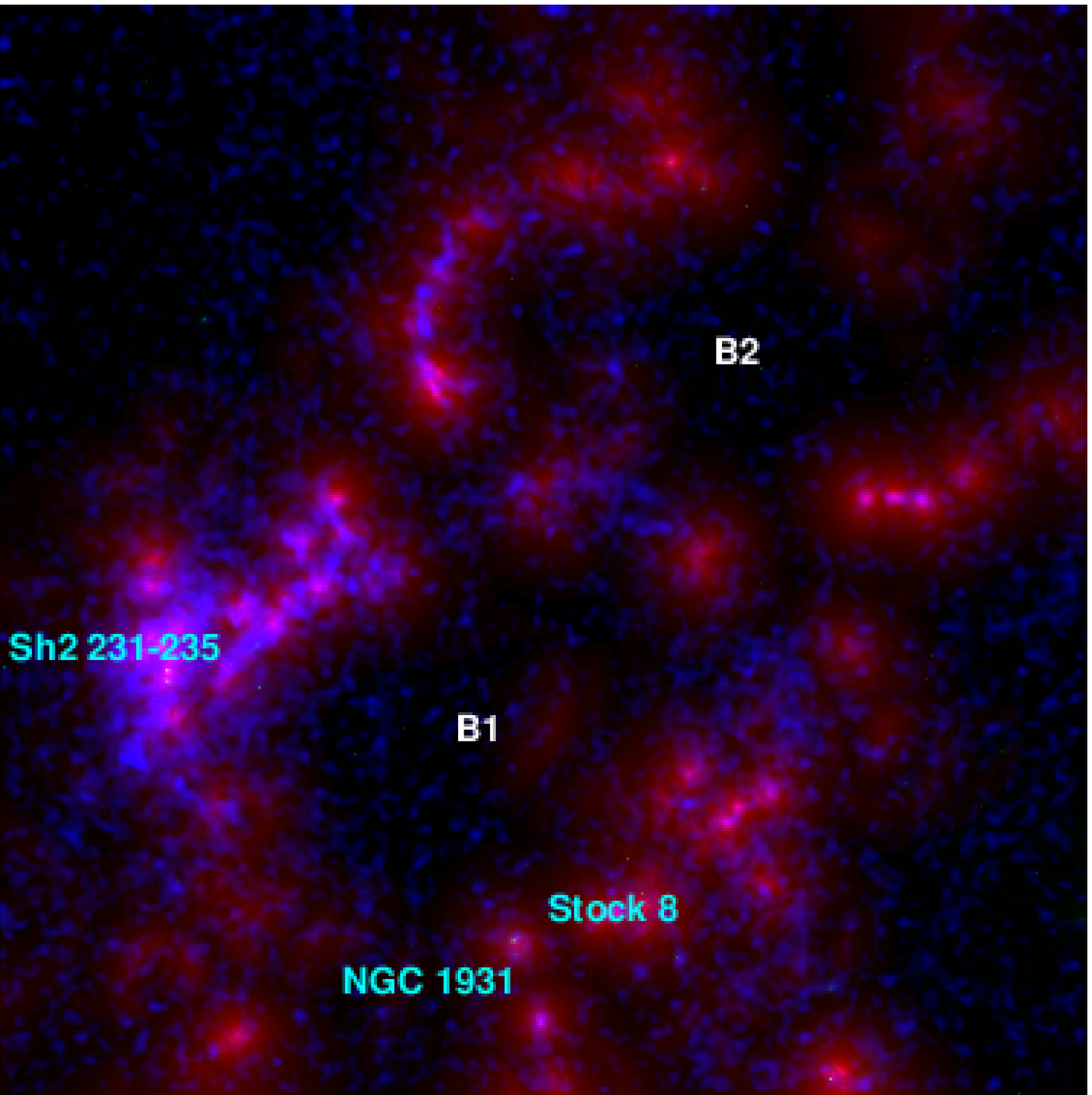}
}
{	
\includegraphics[height=8.5cm,width=9cm]{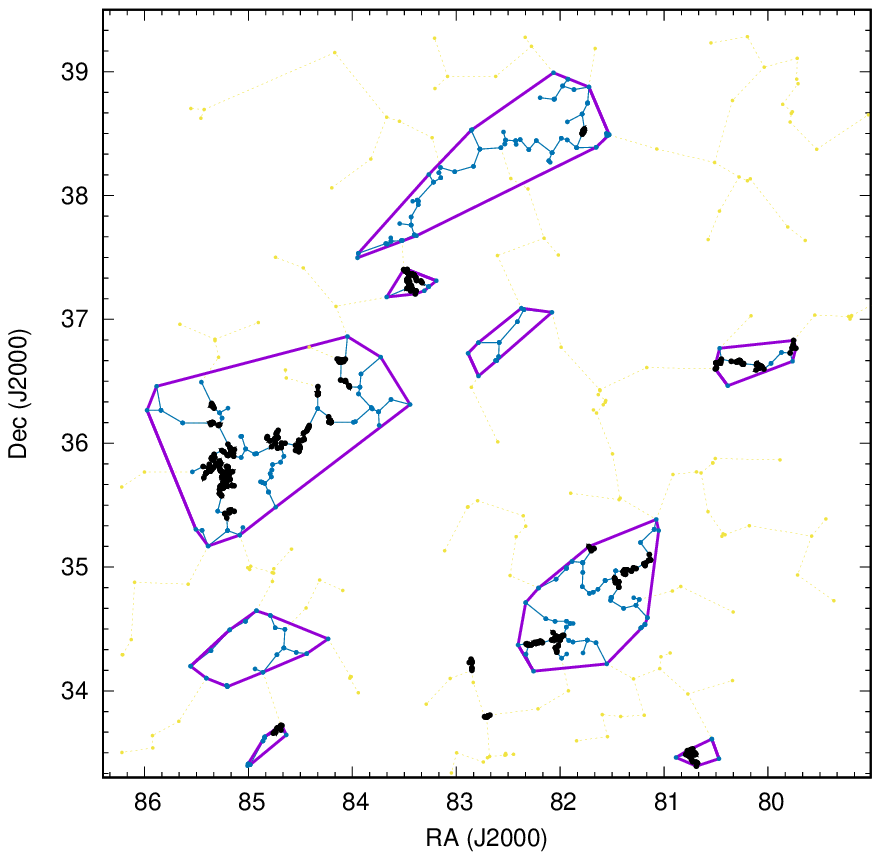}
\includegraphics[height=8.5cm,width=8cm]{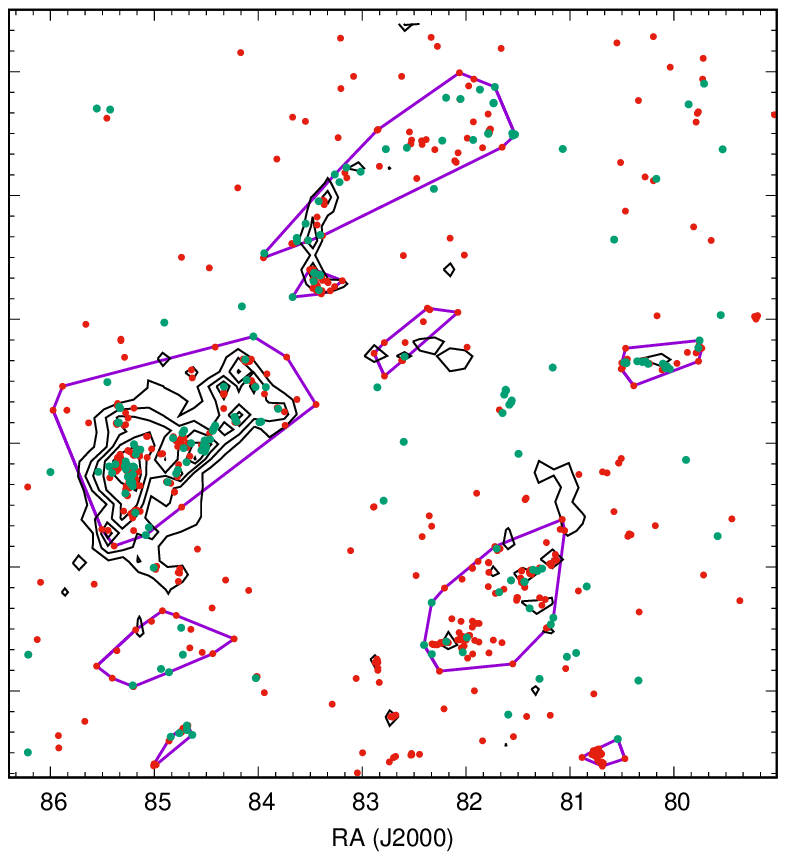}
}
\caption{\label{Fiso} Top left-hand panel: Spatial distribution of YSOs superimposed
on the $6^\circ\times6^\circ$  $WISE$  12 $\mu$m image of the Auriga Bubble region.
Class\,{\sc i} (175) and  Class\,{\sc ii} (535) YSOs are shown by green and red dots, respectively.
The cyan and black curves are the $^{12}$CO and H\,{\sc i} contours taken from \citet{2001ApJ...547..792D} and
\citet{1990A&AS...85..691F}, respectively.
Top right-hand panel: Extinction map (blue colour) and YSOs density map (red colour) 
for the Auriga Bubble region smoothed to a resolution of 18 arcmin.
Bottom left-hand panel: 
MST for the identified YSOs in the Auriga Bubble region
along with the convex hull for active regions.
The black dots connected with solid black lines and the blue dots connected with blue solid 
lines are the branches smaller than the critical length for the sub-clusters and the active regions, respectively. 
The identified active regions are encircled with purple solid lines.
Bottom right-hand panel:  The identified active regions along with the extinction contours (black) and the YSOs distribution.
}
\end{figure*}

\begin{figure*}
\centering
\includegraphics[height=6cm,width=8cm]{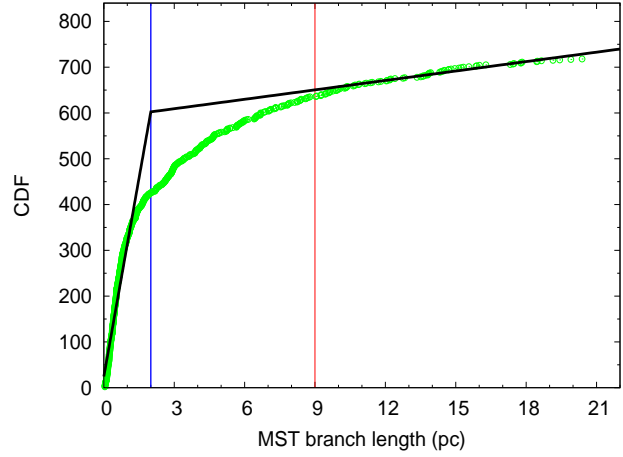}
\caption{\label{Fcdf}
CDF  of MST branch lengths used for the critical length analysis.
The black solid line is a two-line fit to the CDF distribution. 
The inner and outer vertical lines stand for the critical lengths obtained for sub-clusters and active regions, respectively.
 }
\end{figure*}

  Many ground-based NIR surveys of molecular clouds \citep[e.g.,][]{1991ApJ...368..432L,1993ApJ...412..233S, 
2000ApJS..130..381C,2003AJ....126.1916P,2003ARAA..41...57L} have shown that  molecular clouds host both dense `clustered' 
and diffuse `distributed' population.  As discussed earlier the Auriga Bubble region contains several star forming subregions, hence 
a clustered distribution of YSOs is expected.  \citet{2009ApJS..184...18G}  have used an empirical method based on the minimum  
spanning tree (MST) technique to isolate groupings (sub-clusters) from the more diffuse distribution of YSOs in nebulous regions. 
This method  effectively isolates sub-structures without any type of smoothening  \citep[see e.g.,][]{2004MNRAS.348..589C, 
2006A&A...449..151S, 2007MNRAS.379.1302B, 2009MNRAS.392..868B, 2009ApJS..184...18G}. The sub-groups detected in this 
way have no biases regarding  the shapes of the distribution and preserve the underlying geometry of the distribution 
\citep{2009ApJS..184...18G}. In  Fig. \ref{Fiso} (bottom left-hand panel)  we plot the derived MSTs for the YSOs in the region. 
In order to isolate the sub-structures, we adopted a surface density threshold expressed by a critical branch length. 
With the  help of an adopted  MST branch length threshold we can identify local surface density enhancements. To do that, we  used a  
similar approach to that suggested by \citet{2009ApJS..184...18G}. In Fig.  \ref{Fcdf}, we  plot the cumulative distribution for MST 
branch lengths, which shows  a three-segment curve; a steep segment at short lengths, 
a transition segment at the intermediate lengths, and a shallow-sloped segment at long lengths. 
The majority of the sources are found in the steep segment, where the lengths are small 
(i.e., sub-cluster). Therefore, to isolate sub-cluster regions in the Auriga Bubble, we fitted two straight lines to the shallow and steep 
segments of the cumulative distribution function (CDF) and extended them to connect together.  The intersection is adopted  as the 
MST critical branch length, as shown in Fig. \ref{Fcdf} \citep[see also,][]{2009ApJS..184...18G}. 
  The  sub-clusters of the SFRs were then isolated from the lower density distribution by clipping the MST branches longer than the 
critical length described above. Similarly, we defined the  extended area for the SFR by selecting the point 
where the curved transition segment meets the  shallow-sloped segment at longer spacings. 
This range represents the extended region of star formation or the area where YSOs might 
have  moved away from the sub-clusters due to dynamical evolution, and we have named this region as the active region. 
The black dots connected with black lines and the blue dots connected with blue lines in Fig. \ref{Fiso}  (bottom left-hand panel) are 
the branches smaller than the critical length for the sub-clusters and the active region, respectively.
  We have also plotted the convex hulls \citep[cf.][]{2009ApJS..184...18G} for  the active region in Fig. \ref{Fiso} (bottom left-hand 
panel) with solid purple lines. The physical details of the sub-groups (sub-clusters) and the active regions are given in Tables 
\ref{Tp1}, \ref{Tp2} and \ref{Tp3}. In total, we have identified 9 active regions and 26 probable sub-clusters having at-least 5 YSO members in  
the Auriga Bubble region. Although sub-clusters with a small number of members have been reported in  previous studies, e.g., 
\citet[][N=10]{2009ApJS..184...18G}, \citet[][N=7-10]{2004MNRAS.348..589C} and  \citet[][N=10]{2016AJ....151..126S},
however it is worthwhile to mention that smaller numbers of YSOs in some of the probable sub-clusters may introduce relatively large errors in 
the derived physical parameters discussed in the ensuing sub-sections. 
The median value of the critical branch lengths for the sub-clusters and the active regions are 2 pc and 9 pc, respectively.  
In Fig. \ref{Fiso} (bottom right-hand panel) we can see a correlation between the identified active 
  regions with the extinction contours and the YSO locations in the Auriga Bubble.

In many SFRs YSOs are observed to have both diffuse 
and clustered spatial distribution. For example,
\citet{2008ApJ...688.1142K} have analyzed the clustering properties across the W5 region 
and found 40-70\% of the YSOs belong to groups with $\geq$10 members and the remaining were described as scattered 
populations.  Using the AllWISE database, \citet{2016ApJ...827...96F} identified 479 YSOs in a $10\times10$ degree$^2$ region 
centered on the Canis Major star-forming region. Their YSO sample contains 144 and 335 Class\,{\sc i} and Class\,{\sc ii} 
YSOs, respectively. On the basis of the MST of the YSO distribution, \citet{2016ApJ...827...96F} concluded that there were 16 
groups with more than four members. Of the 479 YSOs, 53\% are in such groups.  \citet{2009ApJS..184...18G} presented a 
uniform MIR imaging and photometric survey of 36 nearby young clusters and groups using Spitzer IRAC and MIPS. 
They found 39 clusters/sub-clusters with 10 or more YSO members. Of the 2548 YSOs identified, 1573 (62\%) are
 members of one of these clusters/sub-clusters. Although the  sub-clusters in the Auriga Bubble region have sizes of the order of 
a few parsecs, however, some of the member stars might have moved away in the last few Myr due to dynamical and environmental 
effects \citep{2006MNRAS.369..143M}. \citet{2011MNRAS.410.1861W} used N-body calculations to study the numbers and 
properties of escaping stars from young embedded star clusters during the first 5 Myr of their existence prior to the removal of gas 
from the system. They found that these clusters can lose substantial amounts (up to 20\%) of stars within 5 Myr. In the present 
sample, the YSOs formed in the sub-clusters having a mean velocity of  $\sim$2 km s$^{-1}$ \citep[cf.][]{2011MNRAS.410.1861W} 
can travel a distance of $\sim$2-6 pc in 1-3 Myr of their formation. Therefore, we expect that the effect of escaping members from 
the  sub-clusters/active regions must be insignificant.
We have estimated that the fraction of the scattered YSO population 
(the YSOs outside  sub-clusters, but in the active regions)  is about $\sim$37\% of the total YSOs in 
the whole active regions. Similar numbers ($\sim$40\%) in the case of 8 bright rimmed clouds \citep{2016AJ....151..126S} 
and 5 embedded clusters \citep{2014MNRAS.439.3719C} have also been reported in previous studies. 
The explanation for the scattered populations may include:  escape of sub-clusters/cluster members due to their dynamical 
interaction and isolated star formation \citep[for 
details, ][]{1997ApJ...480..235E,2003ARAA..41...57L,2008ApJ...688.1142K,2009ApJS..181..321E,2014ApJ...787L..15E,2014MNRAS.439.3719C,2019AJ....157..112P}.

\subsubsection{Sub-cluster morphology and structural parameters}

\begin{figure*}
\centering\includegraphics[width=0.75\textwidth,angle=0]{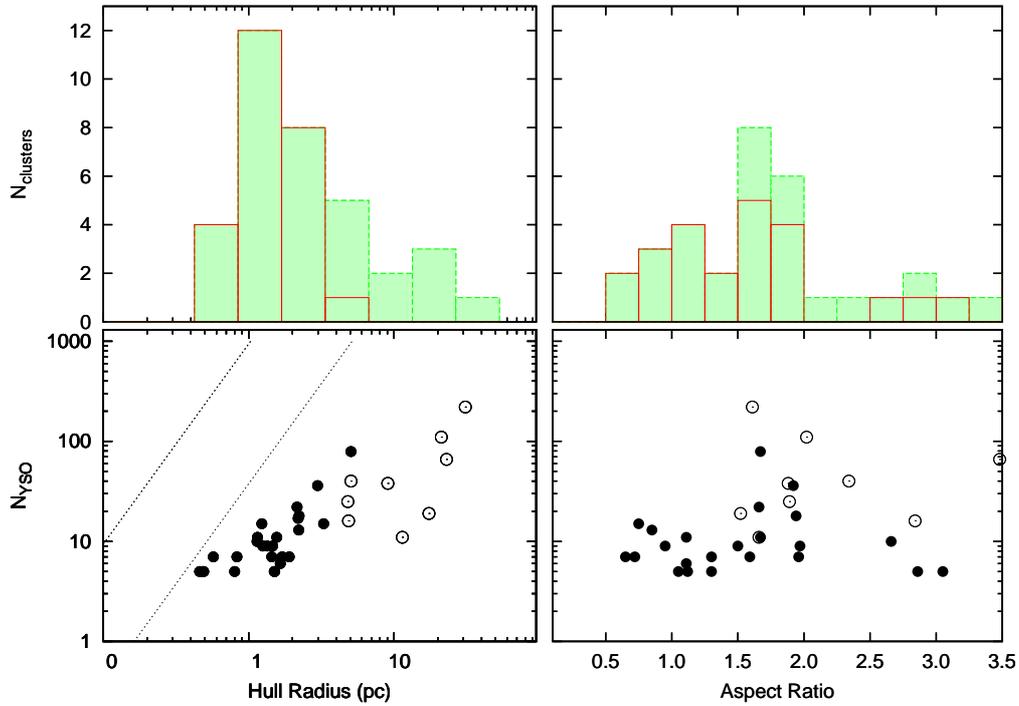}
\caption{\label{Fhull} 
Histogram showing the hull radius distribution (upper left-hand panel) and
plot of the hull radius versus the number of cluster members (lower left-hand panel). 
The red solid histogram and filled circles are for the sub-clusters, and the green dotted histogram
and open circles are for the active regions.
The doted lines in the lower-left panel represent the constant surface densities at 12 and 300 pc$^{-2}$. Those
correspond to the range spanned by the embedded clusters from \citet{2009ApJS..184...18G}.
Right-hand panels: Same as the left-hand, but for the aspect ratio distribution.
 }
\end{figure*}

Known SFRs show a wide range of sizes, morphologies and star numbers 
\citep[cf.][]{2008ApJ...674..336G,2009ApJS..184...18G,2011ApJ...739...84G,2014MNRAS.439.3719C}. We 
use cluster's convex hull radii ($R_H$) and aspect ratios  to investigate their morphology (see Table \ref{Tp2} and Fig. \ref{Fhull}). 
Here we would like to point out that the present procedure is applied to the sample which contains only Class\,{\sc i} and 
Class\,{\sc ii} YSOs and does not include Class\,{\sc iii} YSOs. The similar approach has been adopted in previous studies also 
\citep{2009ApJS..184...18G,2016AJ....151..126S}. However, as discussed in Section 3.2.1, 
the contribution of diskless YSOs (i.e. Class\,{\sc iii} sources) in star-forming regions 
having age 2-3  Myr may be about 50\% of the total YSO population. 
This missing population may play a significant role in estimating 
various parameters discussed in the ensuing sections. Hence we also 
estimate parameters by assuming the contribution of Class\,{\sc iii} sources as 50\% of the total YSO population.

\begin{figure*}
\centering\includegraphics[width=0.75\textwidth,angle=0]{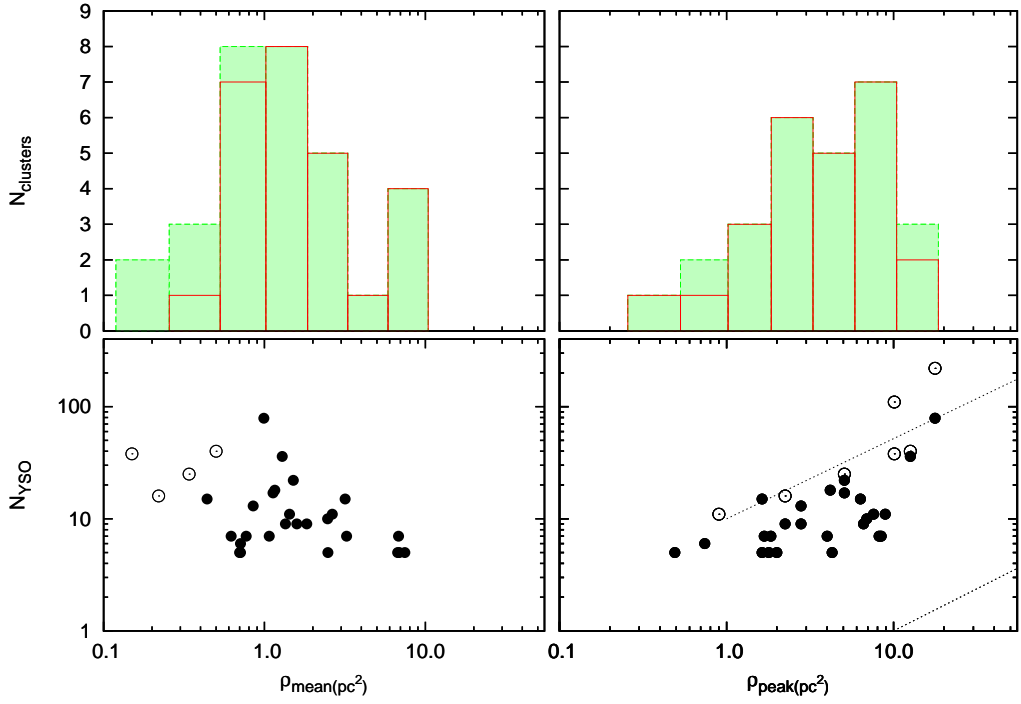}
\caption{\label{Fdensity} 
Histogram showing the  mean YSOs surface density (upper left-hand panel) and plot of 
the mean YSOs surface density versus 
the number of cluster members (lower left-hand panel)
for the sub-clusters and active regions.
The symbols and histograms are the same as Fig. \ref{Fhull}.
Right-hand panels: Same as the left-hand, but for the  peak YSOs surface density distribution.
Dotted lines in the lower-right panel enclose all the regions 
with a slope of 0.8 as given  in \citet{2014MNRAS.439.3719C}.
 }
\end{figure*}

A majority of the  sub-clusters identified in the present sample show an elongated morphology with the median value of  the aspect 
ratios around 1.6. The median number of YSOs in the sub-clusters and the active regions are 9 and 38, respectively 
(cf. Table \ref{Tp3}). The median MST branch length for these  sub-clusters is found to be $\sim$0.5 pc. The total sum of YSOs in 
the active regions  is 546, out of which 345 (63\%) falls in the  sub-clusters. The YSOs in the Auriga Bubble have mean surface 
densities mostly between 0.3 and 7.5 pc$^{-2}$ (see Table \ref{Tp1} and Fig. \ref{Fdensity}). The median values for the surface 
densities for the  sub-clusters and the active region come out to be around 1.35 pc$^{-2}$ and 0.1 pc$^{-2}$, respectively. The peak 
surface densities vary between $\sim$ 0.5 - 18 pc$^{-2}$ with a median value of 4 pc$^{-2}$ for our sample (cf. Table \ref{Tp1} and 
Fig. \ref{Fdensity}). As in the case of low-mass embedded clusters  studied by \citet{2014MNRAS.439.3719C}, we found 
a weak proportionality between the peak surface density and the number of cluster members, suggesting that the clusters are better 
characterized by their peak YSO surface density.

\begin{figure*}
\centering\includegraphics[height=7cm,width=8cm,angle=0]{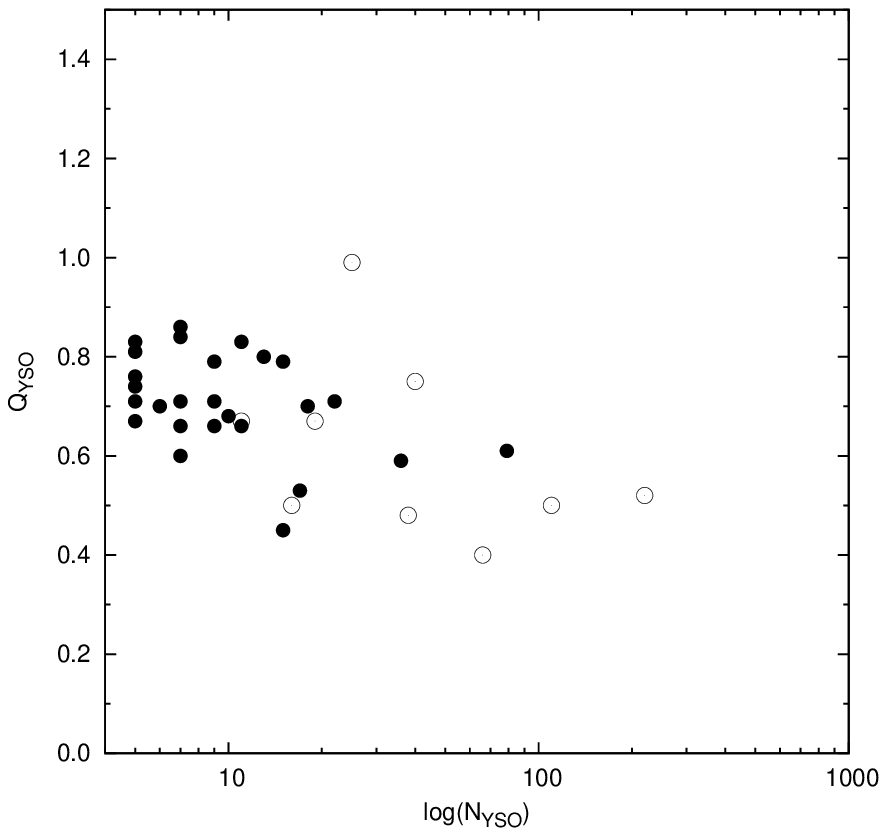}
\centering\includegraphics[height=7cm,width=8cm,angle=0]{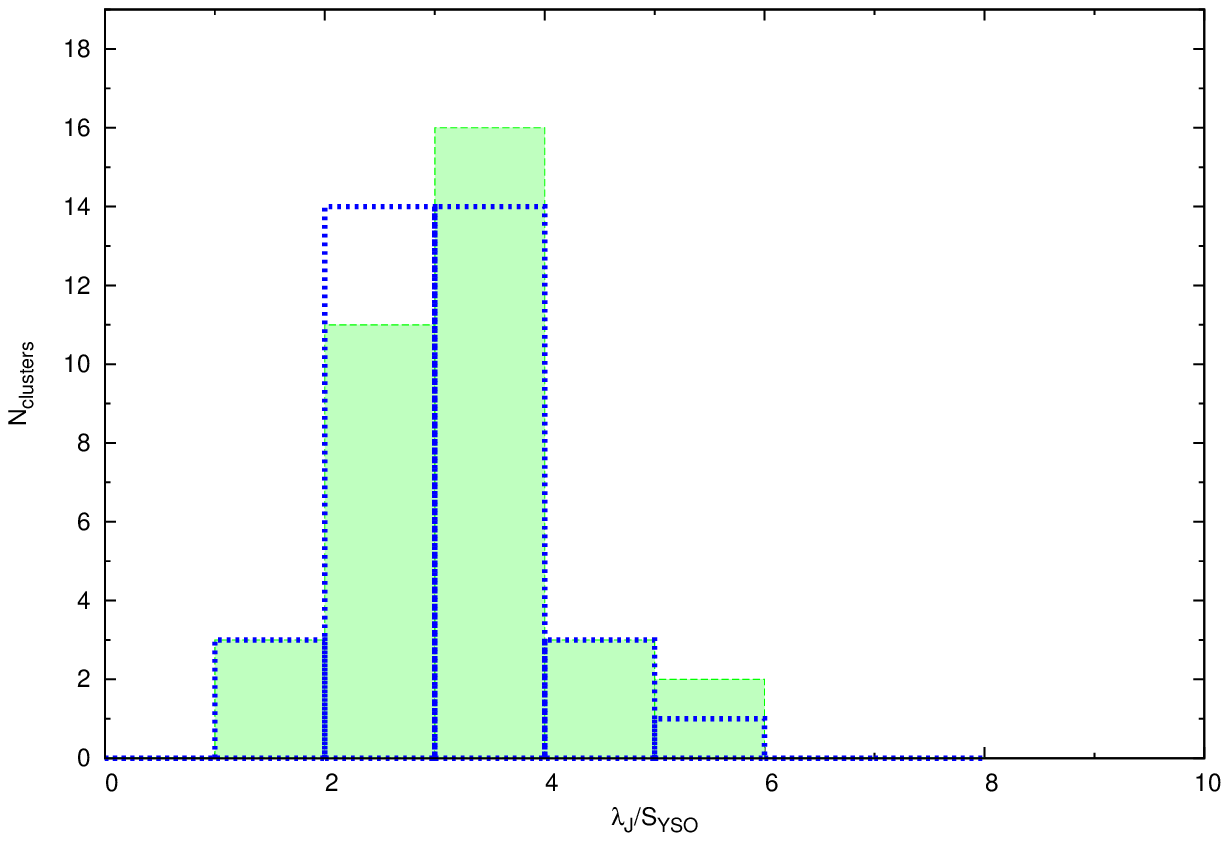}
\caption{\label{Fq} Left-hand panel: Structural $Q$ parameters ($Q$$_{\rm YSO}$) for the sub-clusters 
and active regions. The symbols  are the same as Fig. \ref{Fhull}.
Right-hand panel: Histogram showing the distribution of the ratio between the cluster Jeans 
length ($\lambda_J$) and the mean projected distance between the members of sub-clusters and active regions.}
\end{figure*}

  The spatial distribution of YSOs in a SFR can be investigated with the help of the structural $Q$ parameter 
\citep{2004MNRAS.348..589C,2006A&A...449..151S}. It is used to measure the level of hierarchical versus radial distributions of 
a set of points, and it is defined by the ratio of the MST-normalized mean branch length to the normalized mean separation between 
points \citep[cf.][for details]{2014MNRAS.439.3719C}. If the normal values are used, the $Q$ parameter becomes independent on 
the cluster size \citep{2006A&A...449..151S}. A group of points distributed radially will have a high $Q$ value ($Q$ $>$ 0.8), while 
clusters with a more fractal distribution will have a low $Q$ value ($Q$ $<$ 0.8) \citep{2004MNRAS.348..589C}. 
We find that the groups of the YSOs in the present study (including only Class\,{\sc i} and {\sc ii} sources) have median $Q$ 
values less than 0.8 (0.71 in sub-clusters and 0.52 in active regions, cf. Tables \ref{Tp2} and \ref{Tp3}), indicating a more fractal 
distribution. \citet{2014MNRAS.439.3719C} have found a weak trend in the distribution of $Q$ values per number of members, 
suggesting a higher occurrence of sub-clusters merging in the most massive cluster, which reduces the $Q$ value. A similar trend 
can be noticed in Fig. \ref{Fq}  (left-hand panel).

  \subsubsection{Associated molecular material, stellar mass and Jeans length}

\begin{figure*}
\centering\includegraphics[width=0.75\textwidth,angle=0]{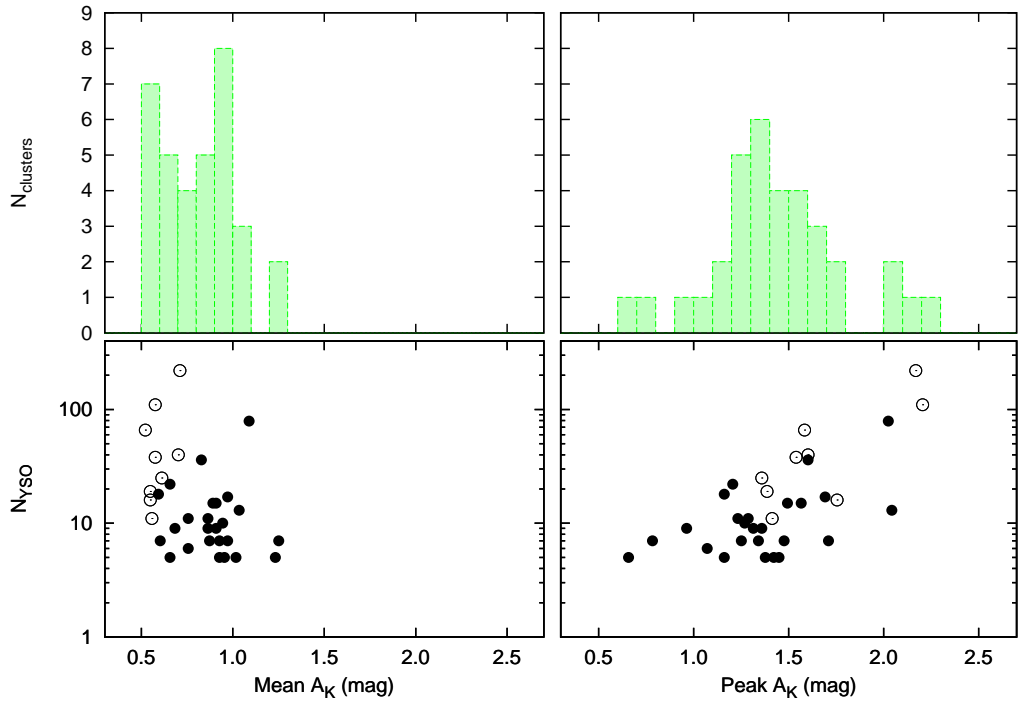}
\caption{\label{Fak} 
Histogram showing the  mean  $K$-band extinction (upper-left panel) and
plot of the mean  $K$-band extinction versus the number of cluster members (lower-left panel). 
The symbols  are the same as Fig. \ref{Fhull}.
Right-hand panels: Same as left-hand, but for the  peak $K$-band extinction distribution.
 }
\end{figure*}

\begin{figure*}
\centering\includegraphics[width=0.45\textwidth,angle=0]{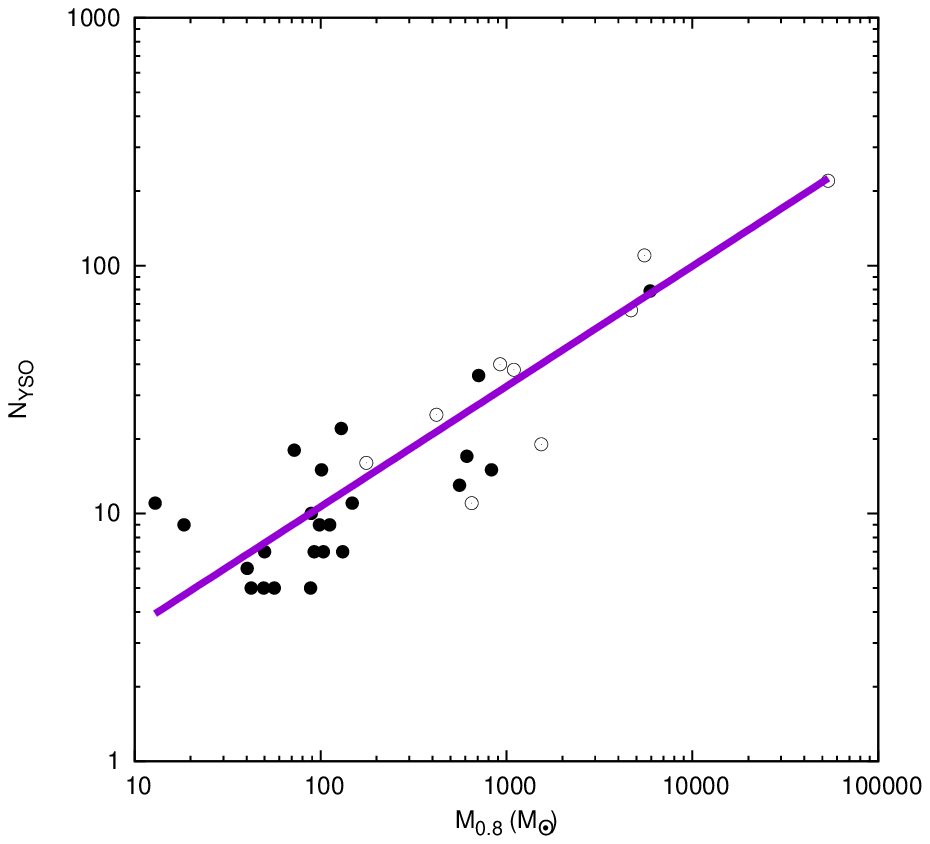}
\centering\includegraphics[width=0.45\textwidth,angle=0]{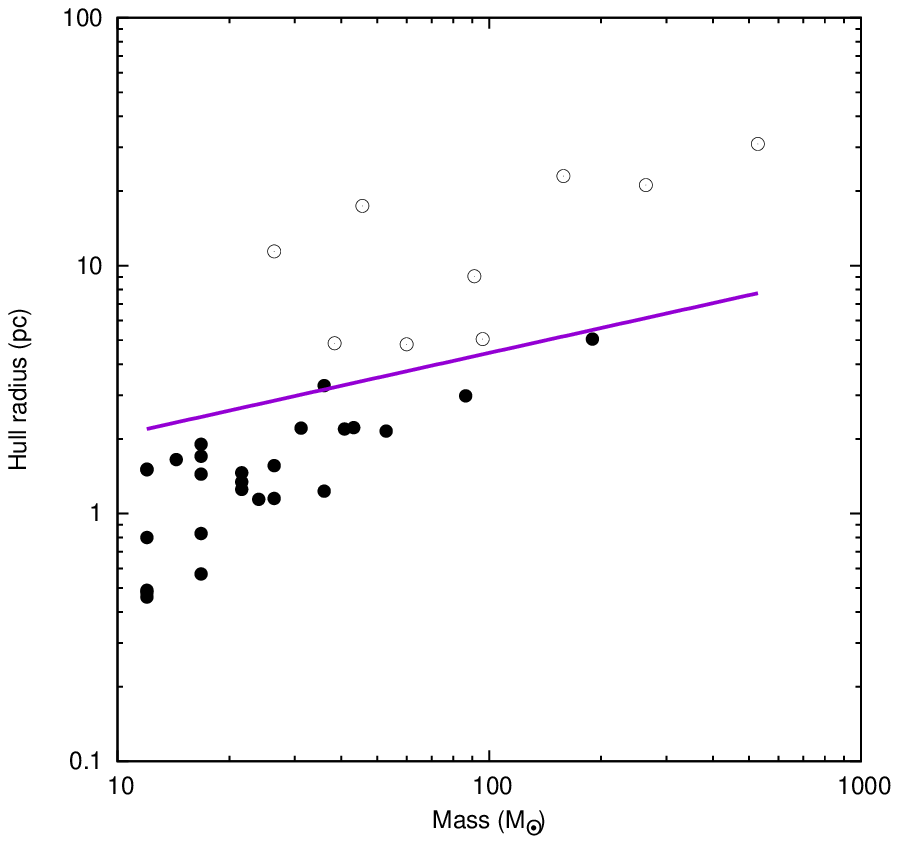}
\caption{\label{Fmass} Left-hand panel: Relation between the number of YSOs `$N_{YSO}$'
and the molecular mass above  $A_K$ = 0.8 mag (M$_{0.8}$) in the sub-clusters and the active regions. 
The number of YSOs is linearly correlated with the cloud mass with Spearman's correlation 
coefficient r=0.8  with 95\% confidence interval of 0.6 to 0.9.  Right-hand panel: Mass of the sub-clusters and the active regions  
are linearly correlated (Spearman's correlation coefficient, r = 0.8  with 95\% confidence interval of 0.6 to 0.9) with the radius of the sub-clusters and the active regions. 
The solid line in right panel represents $r_{lim}$ as a function of the mass (see text). 
The symbols are the same as Fig. \ref{Fhull}.
  }
\end{figure*}

  The  mean $A_K$ values for the identified  sub-clusters have been found in the range of 0.6 and 1.3 mag, with a median value of  
0.9 mag (cf. Table \ref{Tp3} and Fig. \ref{Fak}). 
A weak correlation (Spearman's correlation coefficient `r' = 0.6 with 95\% confidence interval of 0.3 to 0.8) between the 
peak $A_K$ and the number of cluster members can be noticed in Fig. \ref{Fak}.  The median $A_K$ value for the active regions is 
0.6 mag, which is lower than the sub-cluster value, naturally indicating the YSO  distribution of higher density towards the molecular 
clouds of higher density.
The observed correlation indicates that active regions/ sub-clusters having higher number of YSOs have 
higher peak $A_K$ value.
Fig \ref{Fmass} (left panel)  suggests that higher number of YSOs in active regions/ sub-cluster 
are associated with the  higher mass cloud. We presume that massive clouds have higher peak $A_K$ value.

The extinction maps generated in \S 4.2.1 have been used to estimate the molecular mass associated with the identified 
sub-clusters/active regions. The $A_V$ value (corrected for the foreground extinction) in each grid of the 30 arcsec was converted 
to $H_2$ column density by using the relation given by \citet{1978ApJS...37..407D} and \citet{1989ApJ...345..245C},  
i.e., $\rm N(H_2) = 1.25\times10^{21}\times A_V ~cm^{-2}mag^{-1}$. 
The $H_2$ column density was integrated over the convex hull of each region and multiplied by the $H_2$ molecule mass to get 
the cloud mass. The extinction law, $A_K/A_V=0.09$ \citep{1981ApJ...249..481C} has been used to convert $A_K$ values  to 
$A_V$. The foreground contributions have been corrected for by using the relation:  $A_{K_{foreground}}$ =0.15$\times$D \citep[][D 
is distance in kpc]{2005ApJ...619..931I}.  The properties of the molecular clouds associated with the sub-clusters and active regions 
are listed in Table \ref{Tp2}. The  molecular gas associated with the sub-clusters in the present sample shows a wide range in the 
mass distribution ($\sim$5.8 to 7621 M$_\odot$), with a median value around $\sim$146 M$_\odot$.
 SED analysis would have allowed us to estimate the mass of 489 YSOs (cf. Section 3.3.1), 
while for other 221 YSOs available data-points are not enough to properly constrain stellar parameters.
Hence, the total mass of all the candidate YSOs  
in the sub-clusters/ active regions has been estimated  by multiplying the total  number 
of the YSOs (i.e., 710)  with the average SED mass  (2.4 M$_\odot$  as discussed in  Section 4.1).

An analysis of YSO spacings in sub-clusters of 36 star-forming clusters by \citet{2009ApJS..184...18G} suggested that Jeans 
fragmentation is a starting point for understanding the primordial structure in SFRs. We estimated the minimum radius required for 
the gravitational collapse of a homogeneous isothermal sphere (Jeans length `$\lambda_J$') in order to investigate the 
fragmentation scale by using the formula given in \citet{2014MNRAS.439.3719C}. 
  The  Jeans length $\lambda_J$ estimated for the sub-clusters in the current study has values between 0.8 - 3.3  pc
for the sample having Class\,{\sc i} and Class\,{\sc ii} sources as well the sample 
having assumed missing mass of Class\,{\sc iii}. 
The contribution of missing mass of Class\,{\sc iii}  sources has been accounted by assuming a 
disk fraction  of 50\%.
The resultant number of missed stars  has been multiplied with an average SED mass (2.4 M$_\odot$,
as discussed in Section 4.1) to get contribution of unidentified Class\,{\sc iii} sources.
The estimated value of Jeans length `$\lambda_J$'
are given in Tables 9 and 10. The median values of 
`$\lambda_J$' are  estimated as $\sim$2 pc for both of the two samples as discussed. 
We also compared the $\lambda_J$ and the mean separation `$S_{YSO}$' between the cluster 
members (Fig. \ref{Fq}, right-hand panel) and found that the ratio $\lambda_J/S_{YSO}$ has an average value of  $3.4\pm0.9$ 
and $3.2\pm0.8$ for the two samples, respectively. \citet{2014MNRAS.439.3719C} reported the ratio for their sample of 
embedded clusters as  $4.3 \pm 1.5$. The present results agree with a non-thermal driven fragmentation 
since it takes place at scales smaller than the Jeans length \citep{2014MNRAS.439.3719C}.

  \citet{2010ApJ...724..687L} have reported that the number of YSOs in a cluster are linearly related to the dense cloud mass 
M$_{0.8}$ (the mass above a column density equivalent to $A_K \sim$ 0.8 mag) with a slope equal to unity. Recently 
\citet{2014MNRAS.439.3719C} have also found a similar relation for a sample of embedded clusters. This suggests that the star 
formation rates depend linearly on the mass of the dense cloud \citep{2010ApJ...724..687L}. 
We also estimated  the M$_{0.8}$  (cf. Table \ref{Tp2}), accounting only for the Class\,{\sc i} and {\sc ii} objects, and 
found that the number of YSOs in the sub-cluster or active region is linearly correlated with the dense cloud mass with a value of 
Spearman's correlation coefficient `r' = 0.8  with 95\% confidence interval of 0.6 to 0.9 (cf. Fig. \ref{Fmass}, left-hand panel).

The right-hand panel of Fig. \ref{Fmass}  shows the hull radius of sub-clusters/ active regions as a function of the total YSO mass, 
which manifests that the hull radius  is linearly correlated  (r=0.8) to the total YSO mass. 
  The radius versus mass relation gives a clue of whether the cluster will be bound or unbound. The radius limit of a group moving in 
the Galactic tidal field is defined as the distance from the center of the group at which the attraction of a given star from the cluster 
  is balanced by the tidal force of external masses \citep{1969SvA....12..625K}. The limiting radius $r_{lim}$ for a group that is 
moving in elliptical orbit around the Galactic center is given by the relation \citep{1962AJ.....67..471K}.

  \begin{equation}
	  r_{lim} = R_p \Big(\frac{M_*}{3.5 M_G}\Big)^{1/3}
  \end{equation}

  where $R_p$ is the perigalactic distance of the group, $M_*$ is the mass of the group and $M_G$ 
  is the mass of the Galaxy. Assuming $R_p$ of the Sun as 8.5 kpc and $M_G \sim 2\times10^{11} M_\odot$, 
the limiting radius $r_{lim}$ as a function of mass of the cluster is shown as the continuous curve in Fig. \ref{Fmass} (right-hand 
panel). This figure indicates that the radii of all the sub-clusters are below the limiting radius, which suggests that all the sub-clusters 
will be bound systems, whereas all the active regions may be unbound systems.

  \subsubsection{Star formation efficiency}

  The star formation efficiency (SFE), defined as the percentage of the gas mass converted into stars, 
  is an important parameter to determine whether a cluster will be a bound or unbound system etc. Several studies have been 
carried out, suggesting that the formation of a bound system requires a SFE of $\geq$$\sim$ 50\% when the gas dispersal is 
quick or $\geq$$\sim$ 30\% when the gas dispersal time is $\sim$3 Myr \citep[cf.][]{1983MNRAS.203.1011E,1984ApJ...285..141L}. 
  \citet{1984ApJ...285..141L} have also concluded that in the case of slow dispersal a lower  SFE of $\sim$ 15\% may also 
produce a bound system.

  The observed surface density of the YSOs in the sub-clusters and active regions provides an opportunity to study how this quantity 
is related to the observed SFE and other properties of the  associated molecular cloud. Recent works indicate that SFE increases 
with the stellar density; e.g., \citet{2009ApJS..181..321E} reported that YSO clusters of higher surface density have higher SFE 
(30\%) than their lower density surroundings (3\%-6\%). Similarly \citet{2008ApJ...688.1142K} also found SFEs of $>$10\%-17\% 
 for high surface density clusters, whereas in  lower density regions the SFEs are found to be $\sim$ 3\%. 
We have calculated the SFE by using the cloud mass derived from $A_K$ inside the cluster convex hull 
areas and the number of YSOs found in the same areas \citep[see also][]{2008ApJ...688.1142K}. 
{
The total mass of the stellar content was estimated as discussed in Section 4.2.3. 
It is found that for the sub-cluster regions the SFEs varies between 5\% to 20\% with an average of $\sim 10.2\pm 1.2$\%. These 
SFE estimates must be considered as lower limits in these regions as we are considering stellar contents composed of Class\,{\sc i} 
and II sources. Inclusion of Class\,{\sc iii} sources will increase the SFE values.}

\begin{figure*}
\centering\includegraphics[width=0.45\textwidth,angle=0]{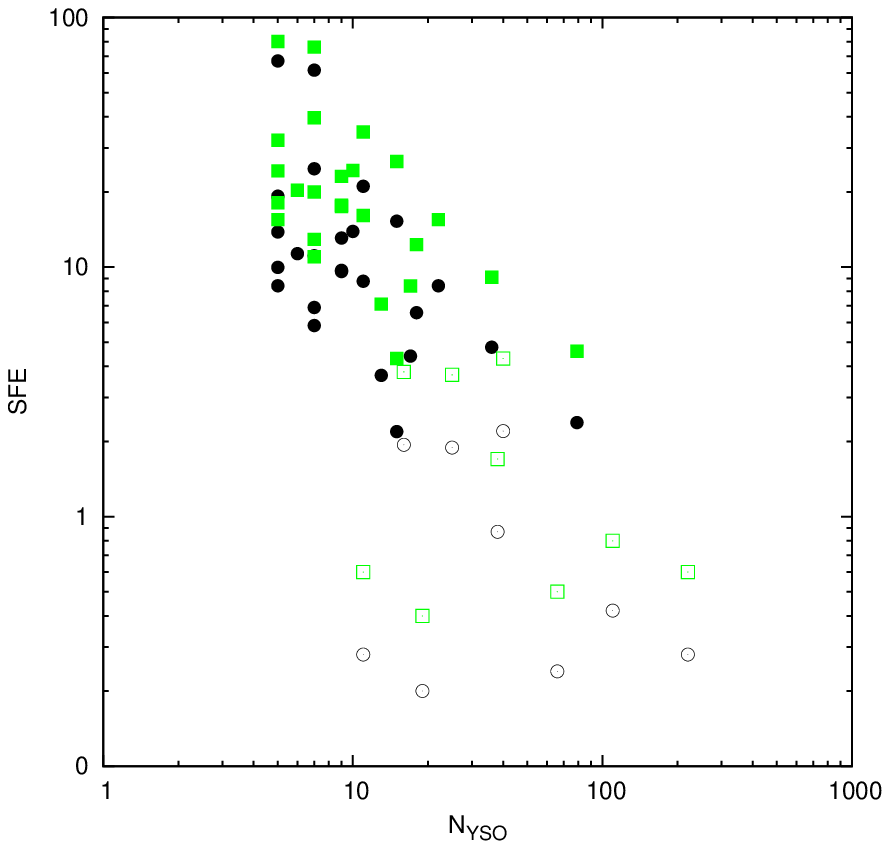}
\centering\includegraphics[width=0.45\textwidth,angle=0]{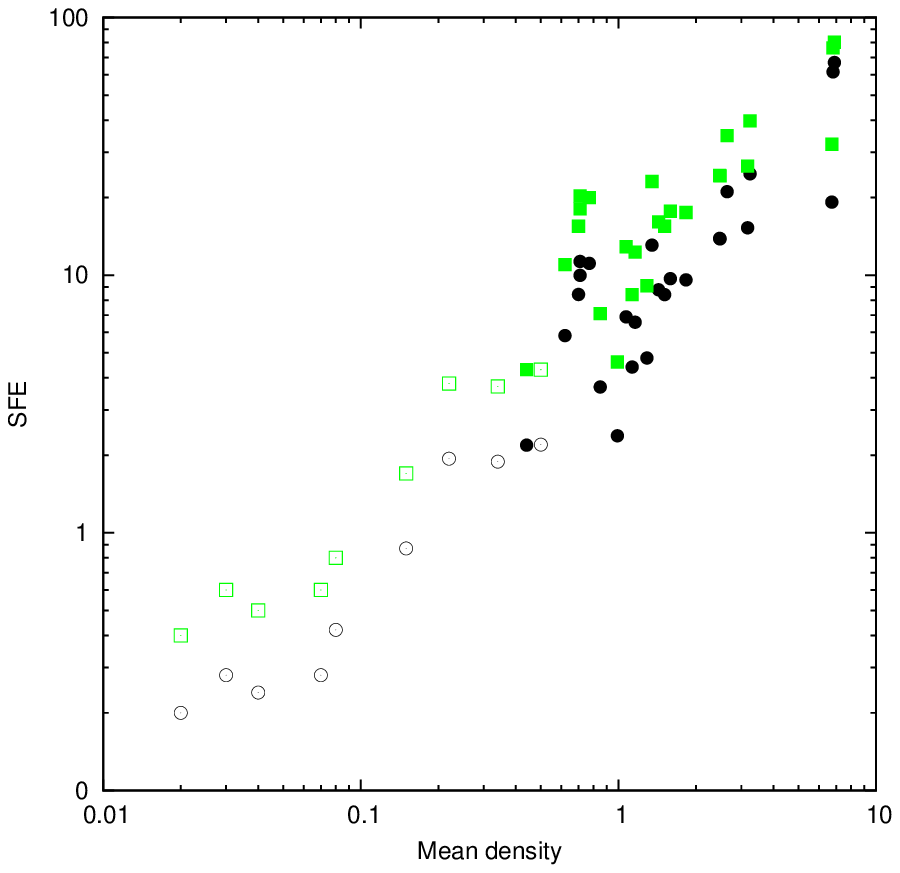}
\caption{\label{Fsfe}  
Star formation efficiency (SFE) (for Class\,{\sc i} and Class\,{\sc ii} YSOs) in the sub-clusters and 
in the active regions with respect to the number of YSOs (left-hand panel) and  
the mean density (right-hand panel). The symbols are the same as Fig. \ref{Fhull}.  
The filled  and open squares represent SFEs with the  assumed contribution of Class\,{\sc iii} sources (see text)  
in the sub-clusters and in the active regions, respectively.
 }
\end{figure*}

  In the case of embedded clusters \citet{2014MNRAS.439.3719C} have obtained SFE as 3-45\% with an average 20\%. 
The SFE distribution as a function of the number of the cluster members and the mean surface density  of each of our regions  is 
shown in Fig. \ref{Fsfe}. The SFE seems to be anti-correlated with the  number of the cluster members 
(Spearman's correlation coefficient `r' =-0.7 with 95\% confidence interval of -0.5 to -0.8) in the sense that the regions associated with high SFEs have smaller numbers of closely packed stars. 
The right-hand panel of Fig. \ref{Fsfe} also indicates a tight correlation between  the SFE and  
mean surface density (Spearman's correlation coefficient `r' =0.9  with 95\% confidence interval of 0.8 to 1.0) 
in the sense that  the regions having higher YSO densities exhibit  higher SFEs. 
However, Fig. \ref{Fsfe} left-hand panel indicates  that the regions associated with high SFEs have smaller numbers of stars. 
The probable explanation for the regions having a smaller number of stars but a higher 
SFE could be that these regions are closely packed as revealed in Fig. \ref{Fdensity}.

\subsection{Star formation history in the Auriga Bubble region: a series of triggered star formation}

\citet{2012AJ....143...75K}  estimated that the shell is expanding with a velocity of  $\sim$55 km s$^{-1}$ and the
 kinetic energy of the shell is $\sim2.5\times10^{50}$ ergs. They also detected hard X-ray emitting hot 
gas inside the shell with a thermal energy of $\sim3\times10^{50}$  ergs. These authors have discussed 
two possibilities for the formation of this shell, i.e., stellar winds from OB stars and 
a SN explosion. However, the wind energy of the eight O stars found in the shell cannot 
explain the large kinetic energy and the hard X-ray emitting hot gas, and so they concluded that a SN created the shell. 

As for the origin of H\,{\sc ii} regions (viz $Sh2~231-235$ and $Sh2~237$) within 
the boundary of the H\,{\sc i}/ continuum structure, \citet{2012AJ....143...75K} discussed that these       
 could have been triggered either by SN explosions, stellar winds, or expanding H\,{\sc ii} regions 
by a previous generation of stars. However, since the ages of the H\,{\sc ii}  regions 
are of the order of a few Myr \citep[see e.g.,][]{2008MNRAS.384.1675J,2013ApJ...764..172P}     
and the estimated age of the  hot shell is only $\sim$0.33 Myr, it is not possible     
that the current expanding shell triggered the formation of these H\,{\sc ii} regions.  
They speculated that the first generation massive stars in the stellar association,
 to which the SN progenitor belonged, could have triggered the formation of the OB stars     
 currently exciting the  H\,{\sc ii} regions. 

In a similar line, we propose the following star formation scenario for the Auriga Bubble1  region. 
The H\,{\sc ii} complex has several O9 type stars \citep[cf. Table 6,][]{2012AJ....143...75K}.
If one or more of the first generation O9 type stars in the association formed on 
an expanding  H\,{\sc ii} regions with a large shell around it,
after $\sim$ 8 Myr \citep[the MS life time of 20 M$_\odot$ O9 type star;][]{2005fost.book.....S}
the dense material collected around it might have collapsed to form the second generation OB stars 
\citep[the collect and collapse mechanism advocated by][]{1977ApJ...214..725E}, 
which are exciting the current H\,{\sc ii} regions (i.e., $Sh2~231-235$ and $Sh2~237$)
 around the shell and their associated clusters. Also in a somewhat similar timescale 
a massive star of the first generation exploded as a SN, forming the shell/Auriga Bubble1. 

Fig. \ref{Fiso} manifests that a majority of the Class\,{\sc i} and Class\,{\sc ii} YSOs identified in the     
 present study are located mostly along the boundary of Bubble1 and Bubble2.  
Their distribution is well correlated with that of the ionized and molecular gas 
as well as PAHs around the H\,{\sc ii} complex G173+1.5.  It is interesting to note that \citet{2008MNRAS.384.1675J}
 noticed several OB stars around the cluster Stock 8 (associated with Sh2 234) and inferred that these 
may belong to the group of the first generation in the region.  They argued that the OB stars within
Stock 8 have ages of 2-3 Myr and were formed by the action of these first generation stars.
 They also noticed a strange `nebulous stream' towards east of Stock 8, which has a group
 of very young YSOs ($\sim$1 Myr) around it, younger than the stars in Stock 8, and argued that the 
star formation in the nebulous stream region is independent and these very young YSOs belong to
 a different generation from those in the Stock 8 cluster. They further inferred that
 these YSOs might have  formed in the remaining clouds due to compression by a  shock front 
from the north. Here it is  worthwhile to mention that the nebulous stream is located 
on the Shell/Bubble1 and north is the direction toward the center of the Shell/Bubble1 
\citep[see Fig. 5 of][]{2012AJ....143...75K}. In addition, a $^{12}$CO cloud is found adjacent just 
to the south of the stream \citep[see Fig. 22 of ][]{2008MNRAS.384.1675J}.

\begin{figure*}
\centering
\includegraphics[width=0.45\textwidth]{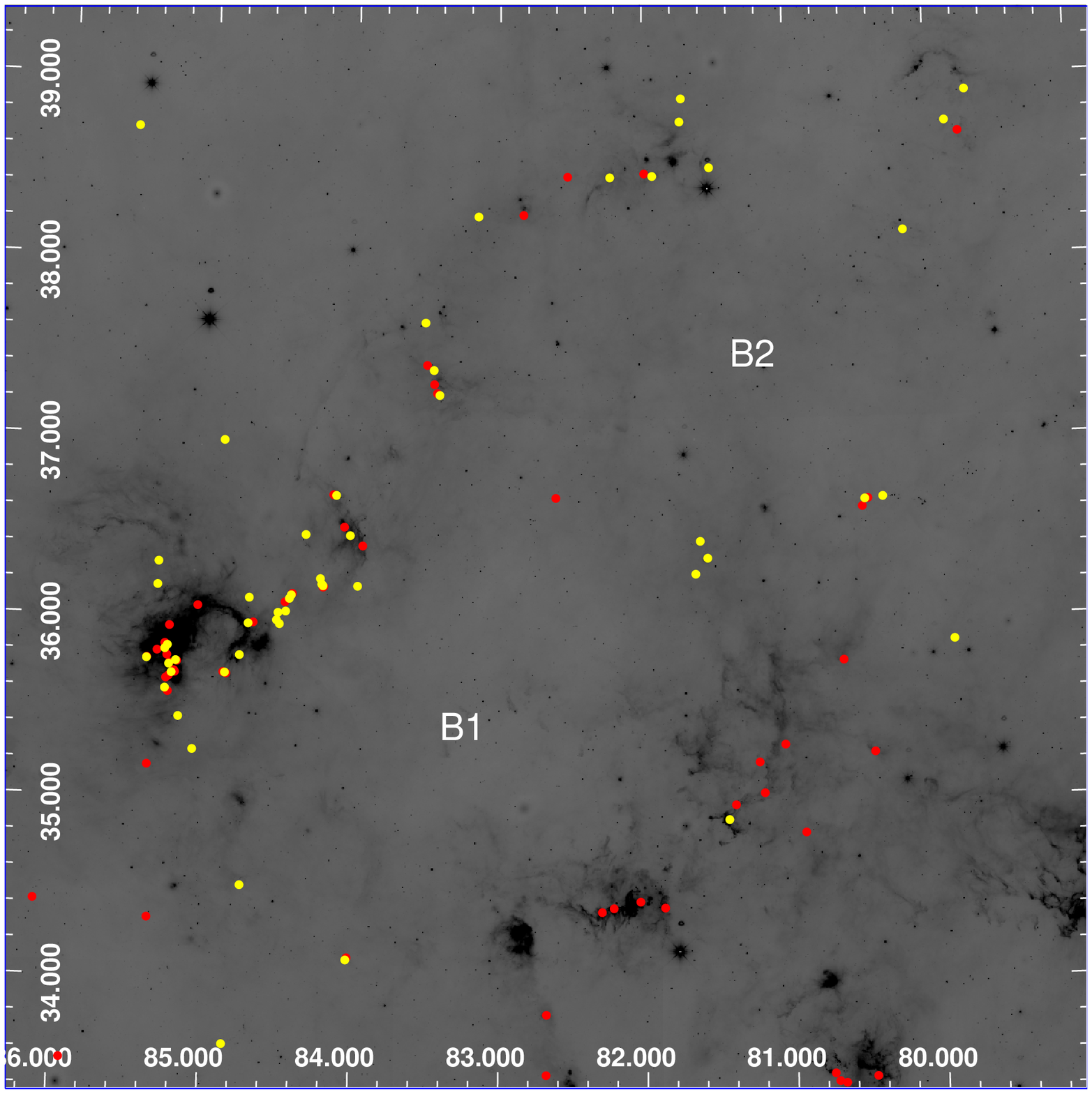}
\includegraphics[width=0.45\textwidth]{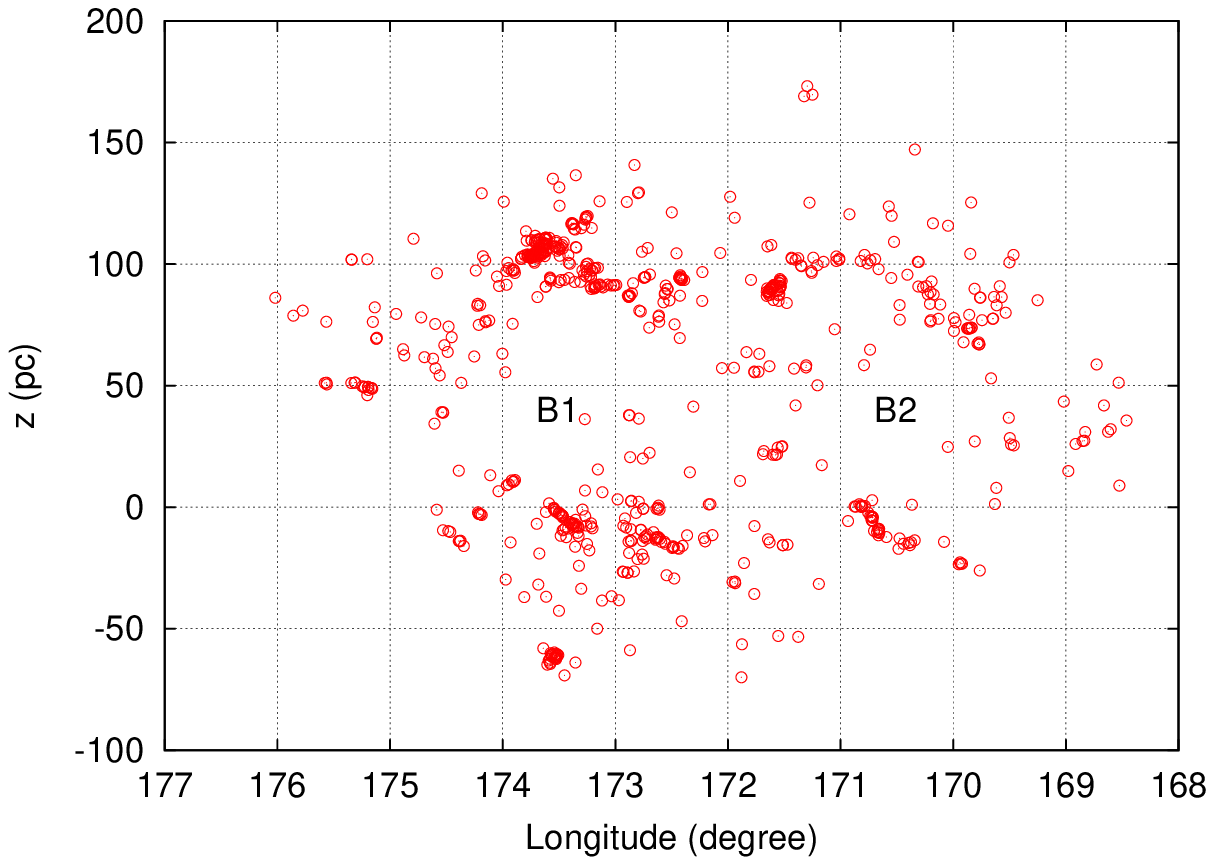}
\caption{\label{gal}  
Left-hand panel: Spatial distribution of the YSOs having age $<$1.0 Myr superimposed
on the  $6^\circ\times6^\circ$  $WISE$  12 $\mu$m image of the Auriga Bubble region.
The X-axis and Y-axis represent RA and DEC at the J2000 epoch.
Right-hand panel: Distribution of all the identified YSOs in the $l-z$ plane.
}
\end{figure*}

Based on this fact we propose that most of the youngest YSOs (having ages $\lesssim$1.0 Myr),  
identified in the present study as well as those in the nebulous stream are of   
 different origin from the somewhat evolved YSO population located in and around the H\,{\sc ii} regions 
   and that they may be a SN- triggered population. Fig. \ref{gal} (left-hand panel) shows the  
   distribution of the YSOs having ages less than 1.0 Myr. The distribution shows that 
  they are located at the periphery of the Bubbles. Keeping in mind the errors in the age  
      estimation and the distribution of the youngest YSOs along the boundary of the H\,{\sc i}  
  and continuum emission, we propose that the expanding Auriga Bubble1 might have  
 compressed the low density molecular material or pre-existing clouds to form a dense
 shell and this became gravitationally unstable to give birth to a new generation of stars. 

Thus the region seems to have a complicated star formation history. 
It seems that the massive stars of the first generation have mostly completed their lives. 
One of them made the SN explosion and created Bubble1. The currently ionizing  sources of the H\,{\sc ii} regions
 located at its periphery (viz $Sh2~231-235$ and $Sh2~237$)
 are the results of the triggered star formation through the collect and collapse process by the first generation population. 
The more or less evolved YSOs around the H\,{\sc ii} regions are probably stars 
of the third generation due to the various star formation activities associated with 
 these  H\,{\sc ii} regions.  However, the majority of the  younger YSOs (having age $\lesssim$1.0 Myr), 
 located at the periphery of Bubble1 can be of different origin from this YSO population, making 
another group of the third generation which was formed due to SN-induced compression presumably.

The size of the Bubble is quite large (diameter $\sim$100 pc) and it could be one of the   
  largest SNR driven bubbles in the Galaxy. \citet{2008ApJ...683..178K} have also reported
 a SNR driven shell in the extreme outer Galaxy having an extent of  $\sim$100 pc,
 which may also have triggered star formation. This shell has survived  for more than 3 Myrs. 

The distribution of the YSOs associated with the Bubble1 and 2 region is shown in Fig. \ref{gal} 
(right-hand panel) on the $l - z$ plane. It is interesting to note that a significant         
 number of YSOs are located above the  Galactic mid-plane. The center of Bubble1 appears to be $\sim$50 pc 
above the Galactic mid-plane. This seems to be consistent with its warping around $l\sim170^\circ - 173^\circ$ 
toward the northern side. The distribution of the blue plume population around $l\sim170^\circ$ in the 
Norma-Cygnus arm also indicates the warping towards the north \citep{2006MNRAS.373..255P}. The            
distribution of the large-scale molecular gas \citep{2006PASJ...58..847N}
 also reveals the northward warp. The model-based distribution of integrated star 
light \citep{2001ApJ...556..181D} and red-clump stars \citep{2002A&A...394..883L}
 also supports the northward warping of the Galactic plane.

  \section{Conclusion}

  Using optical observations and archive NIR and MIR data, we have compiled a catalog of 710 YSOs in an area 
of $6^\circ\times6^\circ$ around the Auriga Bubble region. Of 710 YSOs 175 and 535 are Class\,{\sc i} and Class\,{\sc ii} YSOs, 
respectively. The physical parameters of the YSOs were estimated by using SED model fitting. 
 The spatial distribution of the YSOs and the MIR and radio emission have been used to understand the star formation in 
the region. The followings are the main results:

  \begin{itemize}

	  \item
		  The ages of the majority of the YSOs are found to be $\leq$ 3 Myr, 
		  and the masses are in the range of  $\sim$3-5 M$_\odot$. 
		  Twenty five percent (175 out of 710) of the YSOs are Class\,{\sc i} sources.

	  \item
		  The spatial distribution of the ionized gas as well as the molecular gas is found to be well correlated with that 
		  of the YSOs, which follows two ring-like structures (named as Auriga Bubble 1 and 2) extending over an area 
                   of a few degrees each. The majority of the Class\,{\sc i} sources are found to be distributed along this structure.
                   Auriga Bubble 1 coincides spatially with the high velocity H\,{\sc i} shell discovered by \citet{2012AJ....143...75K}. 

	  \item
		  Twenty six  probable sub-clusters of the YSOs have been identified on the basis of a MST analysis. 
		  The size of the sub-clusters lies in the range of $\sim$0.5 pc to $\sim$ 3 pc. The SFE 
		  and the limiting radius of these sub-clusters suggest that these may be bound stellar groups.

	  \item
		We propose the following possible star formation history in the region: the first generation is an
		already dispersed OB association to which a SN progenitor belonged,
		whereas the second generation is the current H\,{\sc ii} regions and their associated clusters; and the  more or 
		less evolved YSOs distributed in and around  the above H\,{\sc ii} region belong to the third generation. 
		Further there seems to be another, third generation of YSOs, i.e., the youngest population 
		of the region (ages $\lesssim$1 Myr). 
		They are distributed along the periphery of the Bubbles and may have been triggered by the SN 	explosion.

	  \item
		  The center of the bubbles appear to be $\sim$50 pc above the Galactic mid-plane.

  \end{itemize}

  \section*{Acknowledgments}

  We are very thankful to the anonymous referee for the critical review of the contents and useful comments .
  The observations reported in this paper were carried out by using the Schmidt telescope at Kiso Observatory, Japan.  
  We thank the staff members for their assistance during the  observations. 
We are grateful to the DST (India) and JSPS (Japan) for providing financial support to carry out the present study.
  We are also thankful to Dr. Neelam Panwar for critical reading of the manuscript and useful discussions.
  This publication makes use of data products from the Wide-field Infrared Survey Explorer, which is a joint project of the University 
of California, Los Angeles, and the Jet Propulsion Laboratory/California Institute of Technology, funded by the National Aeronautics 
and Space Administration.
  This publication also made use of data from the Two Micron All Sky Survey, which is a joint project of the 
  University of Massachusetts and the Infrared Processing and Analysis Center/California Institute of 
  Technology, funded by the National Aeronautics and Space Administration and the National Science 
  Foundation.

\bibliography{auriga}{}
\bibliographystyle{mnras}

\begin{table*}
\caption{\label{Tlog}  Log of the optical observations with the Kiso Schmidt telescope.}
\begin{tabular}{@{}rrr@{}}
\hline
Date of observations/Filter& Exp. (sec)$\times$ No. of frames & Field\\
\hline
06 December 2012 \\
$V$   &  $180\times3$,$10\times5$ & Auriga 1\\
$I  $ &  $180\times3$,$10\times3$ & Auriga 1\\
$V$   &  $180\times3$,$10\times2$ & Auriga 2\\
$I  $ &  $180\times3$,$10\times4$ & Auriga 2\\
$V$   &  $180\times3$,$10\times4$ & Auriga 3\\
$I  $ &  $180\times3$,$10\times4$ & Auriga 3\\
$V$   &  $180\times3$,$10\times3$ & Auriga 5\\
$I  $ &  $180\times3$,$10\times5$ & Auriga 5\\
\\
10 December 2012 \\
$V$   &  $180\times3$,$10\times3$ & Auriga 4\\
$I  $ &  $180\times3$,$10\times3$ & Auriga 4\\
$V$   &  $180\times3$,$10\times4$ & Auriga 6\\
$I  $ &  $180\times3$,$10\times3$ & Auriga 6\\
\\
27 December 2014 \\
$V$   &  $180\times5$ & Auriga 7\\
$I  $ &  $180\times7$ & Auriga 7\\
$V$   &  $180\times3$ & Auriga 8\\
$I  $ &  $180\times7$ & Auriga 8\\
$V$   &  $180\times10$& Auriga 9\\
$I  $ &  $180\times9$ & Auriga 9\\
\\
\hline
\end{tabular}
\end{table*}

\begin{table*}
\centering
\small
\caption{\label{coeff} 
	Values of the coefficients for calibrating different Auriga fields.}
\begin{tabular}{@{}l@{ }r@{ }r@{ }r@{ }r@{ }r@{ }r@{ }r@{ }r@{}}
\hline

& \multicolumn{4}{c}{Long Exposures} & \multicolumn{4}{c}{Short Exposures}\\
Calibrated Field - Standard Field     & $M1$ & $C1$ & $M2$ & $C2$   & $M1$ & $C1$ & $M2$ & $C2$\\
\hline
Auriga 1 - NGC 1960  &$ 0.013\pm0.002$&$   3.354\pm 0.003 $&$     1.003\pm 0.002  $&$ 0.396\pm 0.002    $&$         0.018 \pm0.002  $&$ 0.208 \pm0.002  $&$    1.032\pm 0.002  $&$ 0.354\pm 0.002 $\\
Auriga 3 - NGC 1960  &$ 0.022\pm0.002$&$   3.346\pm 0.003 $&$     1.029\pm 0.002  $&$ 0.342\pm 0.002    $&$         0.028 \pm0.002  $&$ 0.157 \pm0.002  $&$    1.058\pm 0.002  $&$ 0.305\pm 0.002 $\\
Auriga 2 - Auriga 1  &$ 0.042\pm0.002$&$   3.259\pm 0.003 $&$     0.995\pm 0.004  $&$ 0.261\pm 0.005    $&$         0.028 \pm0.003  $&$ 0.173 \pm0.003  $&$    0.994\pm 0.005  $&$ 0.396\pm 0.004 $\\
Auriga 4 - Auriga 2  &$ 0.046\pm0.001$&$   3.156\pm 0.001 $&$     0.986\pm 0.002  $&$ 0.164\pm 0.003    $&$         0.057 \pm0.002  $&$ 0.143 \pm0.002  $&$    0.990\pm 0.003  $&$ 0.357\pm 0.002 $\\
Auriga 6 - Auriga 4  &$ 0.049\pm0.001$&$   3.306\pm 0.001 $&$     0.984\pm 0.002  $&$ 0.289\pm 0.002    $&$         0.091 \pm0.003  $&$ 0.123 \pm0.002  $&$    1.017\pm 0.003  $&$ 0.241\pm 0.002 $\\
Auriga 5 - Auriga 6  &$ 0.054\pm0.005$&$   2.970\pm 0.006 $&$     0.972\pm 0.003  $&$ 0.120\pm 0.003    $&$         0.056 \pm0.007  $&$ -0.06 \pm0.007  $&$    0.961\pm 0.006  $&$ 0.304\pm 0.004 $\\
Auriga 3 - Auriga 5  &$ 0.083\pm0.004$&$   3.289\pm 0.004 $&$     0.957\pm 0.004  $&$ 0.373\pm 0.003    $&$         0.056 \pm0.005  $&$ 0.095 \pm0.005  $&$    0.967\pm 0.004  $&$ 0.328\pm 0.003 $\\
Auriga 8 - Auriga 4   &$  0.053\pm0.001$&$  3.101\pm0.003$&$    0.978\pm0.002$&$   0.184\pm0.002$&$   -  $&$  -$&$    -$&$   -$\\
Auriga 7 - Auriga 6   &$  0.101\pm0.001$&$  3.194\pm0.001$&$    0.988\pm0.001$&$   0.288\pm0.002$&$   -  $&$  -$&$    -$&$   -$\\
Auriga 9 - Auriga 2   &$  0.053\pm0.002$&$  3.389\pm0.002$&$    0.968\pm0.003$&$   0.652\pm0.002$&$   -  $&$  -$&$    -$&$   -$\\

\hline     
\end{tabular}

$V - v = M1\times(V-I) + C1$\\
$(V-I) = M2\times(v-i) + C2$\\
\end{table*}

\begin{table*}
\centering
\caption{\label{cmpT} 
	 Comparison of the photometric calibration for the Auriga 1 field done with the data from NGC 1960 
	 \citep[][]{2006AJ....132.1669S} and that done with the Auriga 3 data as explained in Section 2.1.
$\Delta$ represents the difference of  NGC 1960 calibration - Auriga 3 calibration.}
\begin{tabular}{@{}rrrr@{}}
\hline
V range& N & $\Delta (V) \pm \sigma$& $\Delta (V-I) \pm \sigma$\\ 
\hline
 9.5-10.5&    69   &$ -0.018\pm  0.020   $&$      0.046\pm  0.036$\\
10.5-11.5&   175   &$ -0.014\pm  0.022   $&$      0.039\pm  0.040$\\
11.5-12.5&   406   &$ -0.006\pm  0.025   $&$      0.025\pm  0.044$\\
12.5-13.5&   867   &$ -0.001\pm  0.024   $&$      0.017\pm  0.043$\\
13.5-14.5&  1584   &$  0.001\pm  0.020   $&$      0.008\pm  0.025$\\
14.5-15.5&  3163   &$  0.008\pm  0.021   $&$     -0.002\pm  0.026$\\
15.5-16.5&  5486   &$  0.016\pm  0.021   $&$     -0.011\pm  0.026$\\
16.5-17.5&  8644   &$  0.023\pm  0.020   $&$     -0.020\pm  0.024$\\
17.5-18.5& 12098   &$  0.032\pm  0.018   $&$     -0.031\pm  0.023$\\
18.5-19.5& 13350   &$  0.043\pm  0.019   $&$     -0.044\pm  0.023$\\
\hline
\end{tabular}
\end{table*}

\begin{table*}
\centering
\caption{\label{cmpT1} Same as Table \ref{cmpT} but for comparison of the photometric calibration of common stars in the 
Auriga 1 and Auriga 3 fields done by using the data of NGC 1960.  }
\begin{tabular}{@{}rrrr@{}}
\hline
V range& N & $\Delta (V) \pm \sigma$& $\Delta (V-I) \pm \sigma$\\
\hline
 9.5-10.5&  12    &$ 0.034 \pm  0.042  $&$ -0.066 \pm 0.038 $\\
10.5-11.5&  31    &$ 0.042 \pm  0.030  $&$ -0.046 \pm 0.031 $\\
11.5-12.5&  92    &$ 0.029 \pm  0.045  $&$ -0.033 \pm 0.061 $\\
12.5-13.5&  182   &$ 0.009 \pm  0.048  $&$ -0.033 \pm 0.071 $\\
13.5-14.5&  290   &$-0.008 \pm  0.037  $&$ -0.011 \pm 0.043 $\\
14.5-15.5&  614   &$-0.026 \pm  0.048  $&$ -0.009 \pm 0.054 $\\
15.5-16.5&  1039  &$-0.039 \pm  0.054  $&$  0.002 \pm 0.058 $\\
16.5-17.5&  1636  &$-0.047 \pm  0.059  $&$  0.012 \pm 0.066 $\\
17.5-18.5&  2317  &$-0.040 \pm  0.067  $&$  0.037 \pm 0.074 $\\
18.5-19.5&  2182  &$-0.027 \pm  0.095  $&$  0.074 \pm 0.108 $\\
\hline
\end{tabular}
\end{table*}

\begin{table*}
\centering
\scriptsize
\caption{\label{data1_yso} Sample of 710 identified YSOs along with their optical (present study), NIR (2MASS) and  MIR ($WISE$) 
bands magnitudes and errors. The complete table is available as online supplementary material. }
\begin{tabular}{@{}c@{ }c@{ }c@{ }c@{ }c@{ }c@{ }c@{ }c@{ }c@{ }c@{ }c@{ }c@{ }c@{}}
\hline
ID                   &  RA     &   Dec    &   $V\pm \sigma$   &   $I\pm \sigma$ &  $J\pm \sigma$   &   $H\pm \sigma$ &   $K\pm \sigma$ &   $W1\pm \sigma$ &   $W2\pm \sigma$ &   $W3\pm \sigma$ &   $W4\pm \sigma$   & Classification$^*$  \\     
\hline
  1 & 79.530584 &+38.373309 &$        -      $&$     -          $&$ 15.507\pm0.080$&$ 14.092\pm   -  $&$ 13.460\pm    -  $& $12.487\pm0.025$&$11.475\pm0.022$&$  8.064\pm0.022$&$ 5.554\pm0.042$&  1 \\ 
  2 & 79.549134 &+37.034571 &$        -      $&$     -          $&$ 16.615\pm0.131$&$ 15.082\pm0.077 $&$ 13.607\pm 0.042 $& $10.999\pm0.024$&$9.826 \pm0.021$&$  7.136\pm0.017$&$ 4.802\pm0.032$&  1 \\ 
  3 & 79.710650 &+38.903351 &$        -      $&$     -          $&$ 16.282\pm0.101$&$ 15.184\pm0.089 $&$ 14.163\pm 0.057 $& $12.366\pm0.023$&$11.287\pm0.022$&$  7.950\pm0.021$&$ 5.164\pm0.031$&  1 \\ 
  4 & 79.753427 &+36.830059 &$        -      $&$     -          $&$ 14.120\pm0.031$&$ 13.081\pm0.033 $&$ 12.348\pm 0.028 $& $11.000\pm0.023$&$9.762 \pm0.021$&$  6.491\pm0.016$&$ 4.228\pm0.022$&  1 \\ 
  5 & 79.765052 &+36.770991 &$        -      $&$     -          $&$ 14.301\pm0.036$&$ 12.597\pm0.033 $&$ 11.018\pm 0.026 $& $9.051 \pm0.022$&$7.530 \pm0.021$&$  4.630\pm0.014$&$ 2.655\pm0.018$&  1 \\ 
\hline
\end{tabular}

$^*$: Classification of YSOs i.e., 1=Class\,{\sc i} (WISE 3 Band), 2=Class\,{\sc ii} (WISE 3 Band),3=Class\,{\sc i} (WISE+2MASS, two Band each),4=Class\,{\sc ii} (WISE+2MASS, two Band each),5=Class\,{\sc i} (from AGN contamination list), 6=Class\,{\sc ii} (WISE 4 BAND, Tr Disk),7=Class\,{\sc i} (2MASS TCD),8=Class\,{\sc ii} (2MASS TCD).\\
\end{table*}

\begin{table*}
\centering
\caption{\label{cftT} Summary of the limiting magnitude and completeness limit for each band.}
\begin{tabular}{@{}cccc@{}}
\hline
Band & Sources in the & Detection  & 90\%  \\
     &  Auriga Region & Limit (mag) & Completeness Limit (mag)\\
\hline
V   &   470876 &  21.21  & 19.00 \\
I   &   470876 &  20.03  & 17.75 \\
J   &   550210 &  17.31  & 15.80 \\
H   &   538281 &  15.51  & 15.10 \\
K   &   500720 &  14.55  & 14.30 \\
W1  &   696985 &  17.06  & 15.75 \\
W2  &   671510 &  17.30  & 15.50 \\
W3  &   50900  &  11.71  & 11.25 \\
W4  &   1001   &   6.80  &  6.25 \\
\hline
\end{tabular}
\end{table*}

\begin{table*}
\centering
\caption{\label{data3_yso}  A sample table containing the stellar parameters of the selected 489 YSOs derived 
by using the SED fitting analysis. IDs are the same as in Table \ref{data1_yso}. 
The complete table is available as online supplementary material. }
\begin{tabular}{@{}lccrrcc@{}}
\hline
   IDs & N  &  Model&$\chi^2$ &  $A_V$    &   Mass    & Age        \\
       &    &       &    &    (mag)   &   (M$_\odot$)& (Mys)  \\            
\hline
   1  & 7   & 501  &$1.8$&$  6.5\pm4.1$  &$1.3\pm1.2$  &$1.2\pm2.2$  \\
   2  & 7   & 213  &$2.7$&$ 12.0\pm4.7$  &$2.9\pm1.1$  &$4.4\pm2.7$  \\
   3  & 7   & 263  &$1.5$&$  3.5\pm3.0$  &$0.9\pm1.0$  &$0.6\pm2.3$  \\
   4  & 7   & 243  &$2.6$&$  6.1\pm3.1$  &$3.2\pm1.6$  &$2.3\pm2.1$  \\
   5  & 7   & 334  &$1.3$&$  8.0\pm3.0$  &$4.6\pm1.6$  &$2.7\pm2.2$  \\
\hline
\end{tabular}
\end{table*}

\begin{table*}
\scriptsize
\centering
\caption{\label{Tp1} Data for identified sub-clusters and active regions. Columns 2 and 3 are their center coordinates.  
The total number of YSOs and their distribution as a function of their evolutionary status are given in column 4 and 
columns 13 to 15, respectively. The hull and circle radius along 
with the aspect ratio are given in columns 6, 7 and 8, respectively. 
Columns 9 and 10 represent the mean and peak stellar density, respectively, obtained by using the isodensity contours. 
Columns 11  and 12 are the mean MST branch length and NN distances, respectively.}
\begin{tabular}{@{}lccrrcccrrrcrrrc@{}}
\hline
Name& $\alpha_{(2000)}$&$\delta_{(2000)}$&N$^a$&V $^b$& $R_{\rm hull}$& $R_{\rm cir}$& Aspect & $\sigma_{\rm mean}$&$\sigma_{\rm peak}$ & MST$^c$ & NN2$^c$ & Class& Class& Frac$^d$ \\
 & {\rm $(^h:^m:^s)$} & {\rm $(^o:^\prime:^{\prime\prime)} $} &  &&  (pc)& (pc)& Ratio & (pc$^{-2}$)& (pc$^{-2}$)  & (pc) &  (pc) &  I& II& (\%) \\
\hline
Sub-clusters\\
\hline
  g0 & 05:27:06.528 & +38:31:00.69&  7& 5 &   0.57&  0.80 & 1.96 & 6.81 &  4.00 &  0.18 & 0.13&  2&  5& 29 \\ 
  g1 & 05:33:44.632 & +37:18:09.48& 36& 7 &   2.98&  4.13 & 1.92 & 1.29 & 12.59 &  0.47 & 0.34&  6& 30& 17 \\
  g2 & 05:36:25.763 & +36:39:57.65& 11& 7 &   1.15&  1.21 & 1.11 & 2.64 &  7.58 &  0.39 & 0.36&  1& 10&  9 \\
  g3 & 05:36:14.706 & +36:29:12.13&  5& 4 &   1.51&  1.72 & 1.30 & 0.70 &  0.49 &  0.41 & 0.22&  2&  3& 40 \\
  g4 & 05:36:51.731 & +36:10:47.44&  7& 5 &   0.83&  0.95 & 1.30 & 3.24 &  8.19 &  0.25 & 0.22&  3&  4& 43 \\
  g5 & 05:37:19.582 & +36:24:39.27&  5& 4 &   0.46&  1.24 & 7.22 & 7.45 &  0.00 &  0.17 & 0.12&  1&  4& 20 \\
  g6 & 05:37:57.437 & +36:01:19.43& 17& 7 &   2.19&  4.37 & 4.00 & 1.13 &  5.07 &  0.68 & 0.41& 12&  5& 71 \\
  g7 & 05:38:56.528 & +36:01:34.84& 15& 7 &   3.28&  2.84 & 0.75 & 0.44 &  1.63 &  1.10 & 0.89&  5& 10& 33 \\
  g8 & 05:40:43.580 & +35:55:49.86& 13& 6 &   2.21&  2.03 & 0.85 & 0.85 &  2.79 &  0.71 & 0.54&  4&  9& 31 \\
  g9 & 05:41:03.392 & +35:44:52.26& 79& 7 &   5.05&  6.52 & 1.67 & 0.99 & 17.70 &  0.48 & 0.28& 18& 61& 23 \\
 g10 & 05:40:43.951 & +35:26:31.58&  7& 5 &   1.90&  1.53 & 0.65 & 0.62 &  1.84 &  0.78 & 0.50&  1&  6& 14 \\
 g11 & 05:41:20.297 & +36:09:24.94&  7& 6 &   1.70&  1.44 & 0.72 & 0.77 &  8.42 &  0.36 & 0.30&  2&  5& 29 \\
 g12 & 05:41:23.041 & +36:18:05.91&  5& 3 &   0.48&  0.81 & 2.86 & 6.89 &  1.63 &  0.47 & 0.30&  2&  3& 40 \\
 g13 & 05:38:51.456 & +33:41:33.88&  9& 6 &   1.46&  1.78 & 1.50 & 1.35 &  2.24 &  0.33 & 0.25&  3&  6& 33 \\
 g14 & 05:24:39.753 & +35:02:36.77&  7& 4 &   1.44&  1.82 & 1.59 & 1.07 &  1.68 &  0.52 & 0.45&  0&  7&  0 \\
 g15 & 05:25:25.517 & +34:58:04.70& 11& 6 &   1.56&  2.02 & 1.67 & 1.43 &  8.91 &  0.53 & 0.22&  3&  8& 27 \\
 g16 & 05:25:49.676 & +34:52:46.42&  5& 4 &   1.50&  1.54 & 1.05 & 0.71 &  4.28 &  0.77 & 0.93&  1&  4& 20 \\
 g17 & 05:26:47.987 & +35:08:53.74&  5& 3 &   0.80&  0.85 & 1.12 & 2.48 &  1.79 &  0.55 & 0.35&  1&  4& 20 \\
 g18 & 05:28:06.715 & +34:24:34.32& 18& 4 &   2.22&  3.09 & 1.94 & 1.16 &  4.17 &  0.64 & 0.41&  2& 16& 11 \\
 g19 & 05:28:57.746 & +34:23:10.62& 15& 8 &   1.23&  2.49 & 4.10 & 3.17 &  6.33 &  0.33 & 0.26&  1& 14&  7 \\
 g20 & 05:30:48.071 & +33:47:43.60&  5& 3 &   0.49&  0.85 & 3.05 & 6.73 &  2.00 &  0.43 & 0.23&  0&  5&  0 \\
 g21 & 05:31:25.302 & +34:13:04.01&  6& 5 &   1.65&  1.73 & 1.11 & 0.71 &  0.74 &  0.75 & 0.60&  0&  6&  0 \\
 g22 & 05:22:55.492 & +33:28:58.03& 22& 7 &   2.15&  2.77 & 1.66 & 1.51 &  5.07 &  0.51 & 0.45&  0& 22&  0 \\
 g23 & 05:21:53.771 & +36:38:38.55&  9& 5 &   1.25&  1.76 & 1.97 & 1.83 &  6.60 &  0.45 & 0.26&  3&  6& 33 \\
 g24 & 05:21:06.340 & +36:39:20.86& 10& 6 &   1.14&  1.85 & 2.66 & 2.47 &  6.90 &  0.37 & 0.26&  4&  6& 40 \\
 g25 & 05:20:20.546 & +36:37:10.68&  9& 5 &   1.34&  1.31 & 0.95 & 1.59 &  2.79 &  0.50 & 0.30&  4&  5& 44 \\

\hline
Active regions\\
\hline
 g26 & 05:30:11.112 & +38:18:03.20& 66&10 &  22.96& 42.82 & 3.48 & 0.04 &  5.24 &  2.63 & 1.19& 26& 40& 39 \\
 g27 & 05:33:42.774 & +37:17:49.50& 40& 6 &   5.05&  7.71 & 2.34 & 0.50 & 12.59 &  0.52 & 0.36&  3& 33&  8 \\
 g28 & 05:39:39.651 & +35:57:23.56&220& 9 &  30.96& 39.23 & 1.61 & 0.07 & 17.70 &  0.69 & 0.49& 64&156& 29 \\
 g29 & 05:39:42.971 & +34:19:10.29& 19& 9 &  17.40& 21.44 & 1.52 & 0.02 &  0.06 &  4.50 & 3.02&  5& 14& 26 \\
 g30 & 05:39:09.134 & +33:37:16.20& 16& 7 &   4.86&  8.19 & 2.84 & 0.22 &  2.24 &  0.77 & 0.33&  5& 11& 31 \\
 g31 & 05:30:12.869 & +36:49:49.87& 11& 6 &  11.41& 14.68 & 1.66 & 0.03 &  0.90 &  4.37 & 1.35&  1& 10&  9 \\
 g32 & 05:26:58.099 & +34:41:14.16&110&10 &  21.13& 30.04 & 2.02 & 0.08 & 10.09 &  1.01 & 0.64& 16& 94& 15 \\
 g33 & 05:22:52.650 & +33:29:09.24& 25& 5 &   4.81&  6.61 & 1.89 & 0.34 &  5.07 &  0.54 & 0.46&  1& 24&  4 \\
 g34 & 05:20:44.558 & +36:39:37.74& 38& 7 &   9.06& 12.42 & 1.88 & 0.15 & 10.09 &  0.72 & 0.45& 13& 25& 34 \\
\hline
\end{tabular}

a: Number of YSOs enclosed in the group;
b: Vertex of the convex hull;
c: Median branch length;
d: Ratio of Class\,{\sc i}/(Class\,{\sc i} + Class\,{\sc ii})
\end{table*}

\begin{table*}
\centering
\caption{\label{Tp2}Properties of the identified sub-clusters and active regions. 
The mean and peak extinction values are given in Columns 2 and 3, respectively. 
Column 4 represents the cloud mass in the convex hull derived by using the extinction map.
Column 5 gives the mass of the dense cloud having $A_K$ greater than 0.8 mag. 
Columns 6/7, 8, 9/10 and 11/12 represent the Jeans length, $Q$ value, SFE, and ratio of the Jeans 
length to the mean separation of YSOs, respectively.}
\begin{tabular}{@{}lrrrrrrrrcrc@{}}
\hline
Name&   $A_{V_{mean}}$ & $A_{V_{peak}}$ &  Mass &    Mass $_{0.8}$ &   \multicolumn{2}{c}{Jeans length (pc)}  & Q & SFE1$^*$  & SFE2$^{**}$  & J/$S_{YSO}$$^*$ &  J/$S_{YSO}$$^{**}$\\
   &     (mag)      & (mag)     & (M$_\odot$) & (M$_\odot$) &  J1$^*$ & J2$^{**}$ &  & && \\
\hline                                                            
Sub-clusters\\
\hline
  g0  &  6.7& 8.7  &    10.3  &     -    & 1.76 &1.48   & 0.60 & 61.5  & 76.2   &  5.6 &4.7 \\
  g1  &  9.2&17.8  &  1688.8  &   708.2  & 2.10 &2.08   & 0.59 &  4.8  &  9.1   &  3.9 &3.9 \\
  g2  &  8.4&14.3  &    96.7  &    12.9  & 2.02 &1.92   & 0.83 & 21.1  & 34.8   &  4.6 &4.4 \\
  g3  & 11.3&15.8  &   127.8  &    88.3  & 2.73 &2.68   & 0.67 &  8.4  & 15.5   &  2.8 &2.8 \\
  g4  & 13.9&16.4  &    50.0  &    50.0  & 1.70 &1.60   & 0.86 & 24.8  & 39.7   &  3.7 &3.5 \\
  g5  &  -  &  -   &     -    &     -    &  -   & -     &   -  &   -   &  -     &   -  & -  \\
  g6  & 10.8&18.8  &   867.2  &   611.4  & 1.85 &1.83   & 0.53 &  4.4  &  8.4   &  2.3 &2.3 \\
  g7  & 10.1&16.6  &  1576.7  &   829.5  & 2.53 &2.52   & 0.79 &  2.2  &  4.3   &  2.4 &2.4 \\
  g8  & 11.5&22.7  &   798.6  &   558.6  & 1.96 &1.94   & 0.80 &  3.7  &  7.1   &  2.5 &2.5 \\
  g9  & 12.1&22.5  &  7621.0  &  5930.7  & 2.20 &2.18   & 0.61 &  2.4  &  4.6   &  4.0 &3.9 \\
 g10  &  9.7&13.9  &   265.7  &   131.6  & 2.69 &2.66   & 0.84 &  5.8  & 11.0   &  3.0 &3.0 \\
 g11  & 10.8&19.0  &   131.6  &    92.4  & 3.20 &3.12   & 0.66 & 11.1  & 20.0   &  5.3 &5.1 \\
 g12  &  7.3& 7.3  &     5.8  &      -   & 1.72 &1.42   & 0.76 & 67.0  & 80.2   &  3.6 &2.9 \\
 g13  &  7.6&10.7  &   140.5  &    18.4  & 2.45 &2.38   & 0.66 & 13.1  & 23.1   &  3.9 &3.7 \\
 g14  & 10.3&14.9  &   222.3  &   103.6  & 1.94 &1.91   & 0.71 &  6.9  & 12.9   &  2.4 &2.4 \\
 g15  &  9.6&13.7  &   268.6  &   148.0  & 1.98 &1.94   & 0.66 &  8.8  & 16.1   &  3.2 &3.1 \\
 g16  & 10.3&15.3  &   106.0  &    56.4  & 2.96 &2.90   & 0.83 & 10.0  & 18.1   &  2.9 &2.9 \\
 g17  & 10.6&12.9  &    73.3  &    42.3  & 1.37 &1.33   & 0.81 & 13.8  & 24.3   &  2.2 &2.2 \\
 g18  &  6.6&12.9  &   601.1  &    72.1  & 2.26 &2.22   & 0.70 &  6.6  & 12.3   &  3.2 &3.1 \\
 g19  &  9.9&17.4  &   195.6  &   101.2  & 1.60 &1.54   & 0.45 & 15.3  & 26.5   &  3.8 &3.7 \\
 g20  & 13.7&16.1  &    49.4  &    49.4  & 0.79 &0.76   & 0.71 & 19.2  & 32.2   &  1.7 &1.6 \\
 g21  &  8.4&11.9  &   110.6  &    40.3  & 3.33 &3.25   & 0.70 & 11.3  & 20.3   &  3.9 &3.8 \\
 g22  &  7.3&13.4  &   563.0  &   129.3  & 2.21 &2.17   & 0.71 &  8.4  & 15.5   &  3.8 &3.7 \\
 g23  & 10.1&15.1  &   199.3  &    98.5  & 1.64 &1.61   & 0.71 &  9.6  & 17.5   &  2.8 &2.7 \\
 g24  & 10.5&14.1  &   145.7  &    89.1  & 1.66 &1.61   & 0.68 & 13.9  & 24.4   &  3.3 &3.2 \\
 g25  &  9.6&14.6  &   196.8  &   111.8  & 1.84 &1.80   & 0.79 &  9.7  & 17.7   &  3.0 &2.9 \\
\hline               
Active regions\\     
\hline               
 g26  &  5.8&17.6  & 64530.5  &  4674.1  & 7.35 &7.35   & 0.40 &  0.2  &  0.5   &  2.6 &2.6 \\
 g27  &  7.8&17.8  &  4185.3  &   925.2  & 2.96 &2.95   & 0.75 &  2.2  &  4.3   &  3.7 &3.7 \\
 g28  &  7.9&24.1  &182462.2  & 53738.4  & 6.84 &6.84   & 0.52 &  0.3  &  0.6   &  4.7 &4.7 \\
 g29  &  6.1&15.4  & 22525.6  &  1537.7  & 8.21 &8.20   & 0.67 &  0.2  &  0.4   &  1.7 &1.7 \\
 g30  &  6.1&19.5  &  1905.0  &   176.1  & 4.15 &4.13   & 0.50 &  1.9  &  3.8   &  2.9 &2.9 \\
 g31  &  6.2&15.7  & 9304.3   &   649.5  & 6.78 &6.78   & 0.67 &  0.3  &  0.6   &  1.5 &1.5 \\
 g32  &  6.4&24.5  & 61706.7  &  5518.6  & 6.63 &6.63   & 0.50 &  0.4  &  0.8   &  3.8 &3.8 \\
 g33  &  6.8&15.1  &  3047.0  &   419.7  & 3.23 &3.22   & 0.99 &  1.9  &  3.7   &  2.7 &2.7 \\
 g34  &  6.4&17.1  & 10179.2  &  1095.2  & 4.58 &4.57   & 0.48 &  0.9  &  1.7   &  3.5 &3.5 \\
\hline
\end{tabular}

$^*$: For Class\,{\sc i} and Class\,{\sc ii} sources only.

$^{**}$: With inclusion of assumed missing contribution of  Class\,{\sc iii} sources.

\end{table*}

\begin{table*}
\centering
\caption{\label{Tp3} Median value of the parameters of all the  sub-clusters and active regions. 
}
\begin{tabular}{@{}lll@{}}
\hline
Properties  & Sub-cluster   &  Active region\\
\hline
Number of YSOs                 &  9      & 38     \\
R$_{\rm hull}$  (pc)           &    1.5 &   11.4  \\
Aspect Ratio                   &    1.6 &   1.9  \\
Mean number density (pc$^{-2}$)&   1.4    &  0.1     \\
Peak number density (pc$^{-2}$)&  4    & 5.1     \\
A$_K$ (mag)               &    0.9  &   0.6  \\
Peak  A$_K$(mag)               &    1.3  &   1.6  \\
Cloud mass (M$_\odot$)         &  146    & 10179     \\
Dense cloud mass (M$_\odot$)   &  99    & 1095     \\
MST branch length (pc)         &    0.47 &   0.77  \\
Structural $Q$ parameter         &    0.71 &   0.52  \\
Jeans Length (pc)              &    2.0 (1.9)$^*$ &   6.6 (6.6)$^*$  \\
Star formation efficiency (\%) &     9.7 (17.7)$^*$  &    0.4 (0.8)$^*$ \\
\hline
\end{tabular}

$^*$: Values within parenthesis are the estimates with inclusion of assumed contribution of Class\, {\sc iii} sources.
\end{table*}

\clearpage

\bsp    
\label{lastpage}
\end{document}